\documentclass[aps,prd,twocolumns,eqsecnum,preprintnumbers,showpacs,amsmath,amssymb]{revtex4}

\usepackage{graphicx}
\usepackage{dcolumn}
\usepackage{bm}

\begin{document}

\title{ TRUE DYNAMICAL AND GAUGE STRUCTURES OF THE
QCD GROUND STATE AND THE MASSIVE GLUON FIELDS}

\author{Gergely G\'abor Barnaf\"oldi}
\email[]{barnafoldi.gergely@wigner.hu}
\author{Vakhtang Gogokhia}
\email[]{gogohia.vahtang@wigner.hu}

\affiliation{Wigner Research Centre for Physics, 29-33 Konkoly-Thege Mikl\'os Str, H-1121, Budapest, Hungary}

\date{\today}

\begin{abstract}
We convincingly argue that the true dynamical and gauge structures of the QCD ground state are much more complicated than its Lagrangian
exact gauge symmetry supposes to be.
The dynamical source of these complications has been identified with the tadpole/seagull term, which renormalized version called, {\sl the mass gap}.
It is explicitly present in the full gluon self-energy. The true dynamical role of the mass gap was hidden in the QCD ground state.
To disclose it the splintering between the transverse conditions for the full gluon self-energy and its subtracted counterpart has been derived.
We have established the equation of motion for the full gluon propagator on account of the mass gap. It allows to fix the dynamical and gauge structures
of the full gluon propagator with a newly-derived generalized gauge.
It is uniquely given as a function of the gluon momentum. All this makes it possible to present novel non-perturbative analytical approach
to QCD, which we call the mass gap approach. It extends the mass gap concept to be accounted for the QCD ground state as well.
We have found the general non-perturbative solution for the full gluon propagator with exactly defined finite
gluon pole mass, and it is different from the excitations with the effective gluon masses.
Despite the exact gauge symmetry is broken in the ground state, the renormalizability of the theory is not affected due to the important role of the
Slavnov-Taylor identity in the renormalization beyond the perturbation theory.
For this the non-perturbative multiplicative renormalization program for the full massive gluon propagator has been formulated.
Our approach does not allow the massive gluons to be the mass-shell objects, and
thus prevents them to appear as physical states at large distances (confinement of massive gluons).
\end{abstract}



\maketitle

\section{Introduction}

Quantum Chromodynamics (QCD) is widely accepted as the well-functioning, quantum field gauge theory of strong interactions, which works not only at the fundamental quark-gluon level, but at the more complex hadronic level as well~\cite{1,2,3,4,5,6,7,65}. This theory should describe the properties of the observed hadrons in terms of the non-observable quarks and gluons from first principles.
In parallel, it is to be complemented by the quark model (QM), which treats the strongly-interacting particles (baryons and mesons)
as bound-states of quarks, emitting and absorbing gluons. This purpose remains a formidable task yet because of the multiple
dynamical and topological complexities of low-energy particle physics, originated from QCD and its ground state.
This happens because QCD as a fundamental quantum field gauge theory still suffers from a few important conceptual problems.
We focus on some of them as follows:

\begin{itemize}
\item[(A)] The dynamical generation of a mass squared at the fundamental quark-gluon level, since
the QCD Lagrangian forbids such kind of terms apart from the current quark mass.
\item[(B)] Whether the symmetries of the QCD Lagrangian and its ground state coincide or not?
\item[(C)] The extension of the mass gap concept to be accounted  for the QCD ground state, since first it has been defined
within the Hamiltonian formalism.
\item[(D)] The non-observation of the colored objects as physical states which does not follow from the
QCD Lagrangian, i.e., it cannot explain confinement of gluons and quarks.
\end{itemize}

The properties and symmetries of the QCD Lagrangian, and thus including its Yang-Mills (YM) part, are
well-known~\cite{1,2,3,4,5,6,7,8,65} (and references therein).
As mentioned above, any mass scale parameter apart from the current quark mass, explicitly violates the $SU(3)$ color gauge invariance/symmetry of the QCD Lagrangian, for example such as the massive gluon term $M^2_gA_{\mu}A_{\mu}$. It can be treated as the mass gap, the concept introduced first by
Jaffe and Witten (JW)~\cite{9} within the framework of the QCD/YM Hamiltonian formalism. The important aim of our investigation here is to extend the concept of the mass gap to be accounted for the QCD/YM ground state (vacuum) as well. Therefore, we must address to the system of dynamical equations of motion, the so-called Schwinger\,--\,Dyson (SD) system of equations, describing the interactions and propagations of quarks and gluons in the QCD vacuum~\cite{2,8,10,11,12,13,14,15}.
Thus, if there is no room in the QCD Lagrangian  for the mass scale parameter(s), then the only place where it may
explicitly appear is this system, indeed, which contains the full dynamical information on QCD.
In other words, it is not enough to know the Lagrangian of the theory, but it is also necessary and important to know
the true dynamical and gauge structures of its ground state. Furthermore, there might be symmetries of the Lagrangian which do not coincide with symmetries of the vacuum and vice versa. These equations should be also complemented by the corresponding Slavnov\,--\,Taylor (ST) identities, which connect lower- and higher Green's functions (propagators and vertices) to each other~\cite{2,8,10,11,12,13,14,15,16,17,18,19,20,21,22,65}. These identities are consequences of the exact gauge invariance and are important for renormalizability of the QCD. They {\sl "are \ exact \ constraints \ on \ any \ solution \ to \ QCD"}~\cite{2}.
The SD system of dynamical equations, complemented by the ST identities, can serve as an adequate and effective tool for the non-perturbative (NP) analytical
approach to low-energy QCD. Their investigation may reveal much more dynamical information on the QCD ground state, than its Lagrangian may provide at all.
In connection to this let us note that
there exists another powerful NP approach--the lattice QCD--to calculate the properties of low-energy particle physics~\cite{23}.
We believe that these two NP approaches should complement each other, in order to increase our understanding of quantum field theories of particle physics.

We organize our paper as follows. In Section II we introduce the gluon SD equation. In Section III the transversity of the full gluon self-energy
is investigated without any use of the perturbation theory (PT). We discuss under which conditions the exact gauge symmetry might be preserved in the QCD ground state. Then we have shown that the exact gauge symmetry of the QCD Lagrangian is dynamically broken in its ground state.
All this made it possible to extend the mass gap concept to be accounted for the the QCD vacuum as well.
In Section IV the mass gap approach to QCD is formulated within a newly-derived generalized gauge, following by the NP renormalization of the full massive gluon propagator in Section V. In Section VI the asymptotical properties of the full massive solution have been analysed.
In Section VII the full massive gluon propagator in the canonical gauge is discussed. In Section VIII the dynamical and gauge structures of the full massive
gluon propagator on the mass-shell have been investigated in detail. We discuss and summarize our results in Section IX. In Appendix A we have demonstrated  the uniqueness of our approach. We provide the results in Euclidian metric for lattice theory comparison in Appendix B.

\section{The gluon SD equation}

The propagation of gluons is one of the main dynamical effects in the QCD vacuum.
The importance of the corresponding equation of motion is due to the fact that its solutions are supposed to reflect the quantum-dynamical structure of the QCD ground state. The gluon SD equation is a highly non-linear (NL) one because of the self-interaction of massless gluon modes, so the number of its independent solutions is not fixed {\sl a priori}. These solutions have to be considered equally. It is important to underline in advance that unlike Quantum
Electrodynamics (QED)~\cite{24}, in QCD any deviation of the full gluon propagator from the free one requires the presence of the mass squared scale parameter
on the general dimensional ground (see below).
The structure of the gluon SD equation is present in this section in some necessary details. For our purpose it is convenient
to begin with the general description of the SD equation for the full gluon propagator $D_{\mu\nu}(q)$. Analytically it can be written down as follows:

\begin{equation}
D_{\mu\nu}(q) = D^0_{\mu\nu}(q) + D^0_{\mu\rho}(q) i \Pi_{\rho\sigma}(q; D) D_{\sigma\nu}(q),
\end{equation}
where $D^0_{\mu\nu}(q)$ denotes the free gluon propagator, while $\Pi_{\rho\sigma}(q; D)$ is the full gluon self-energy which depends on the full gluon propagator due to the non-abelian character of QCD.
Here and everywhere below we omit the color group indices, since for the gluon propagator (and hence for its self-energy) they factorize, for example $D^{ab}_{\mu\nu}(q) = D_{\mu\nu}(q)\delta^{ab}$. The gluon SD equation (2.1) in terms of the corresponding skeleton loop diagrams is shown below in Fig. 1.

\begin{figure}[h!]
\begin{center}
\includegraphics[width=13.0truecm]{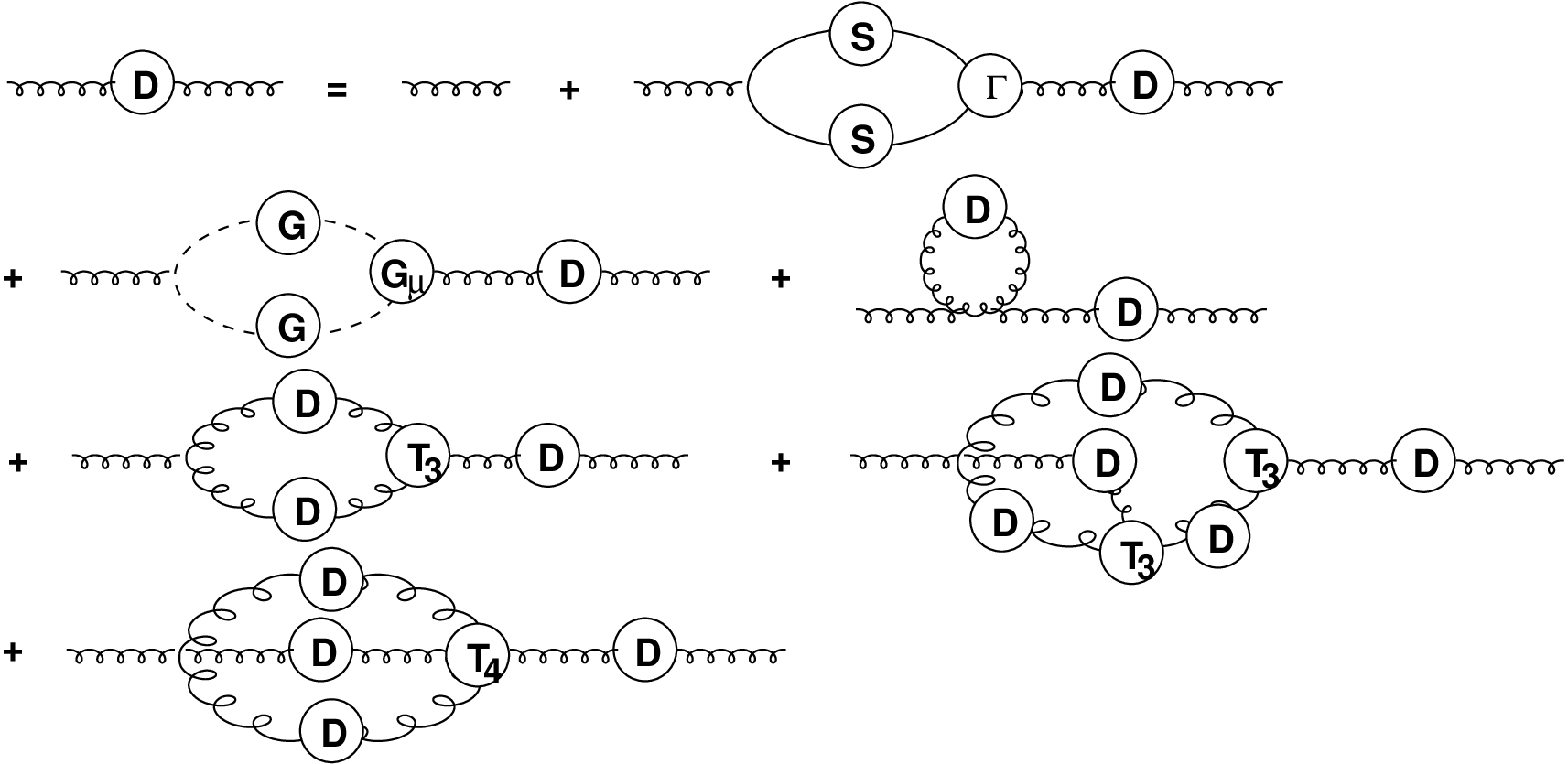}
\caption{The SD equation for the full gluon propagator as present in Ref.~\cite{8}.}
\label{fig:1}
\end{center}
\end{figure}

Here helix/stringy line is for the free gluon propagator, while $D$ denotes its full counterpart. The $S$ with solid lines denotes the full quark propagator, and $\Gamma$ denotes the full quark-gluon vertex. $G$ with dashed lines denotes the full ghost propagator, and $G_{\mu}$ is
the full ghost-gluon vertex.
Finally, $T_3$ and $T_4$ denote the full three and four-gluon vertices, respectively.
Fig. 1 shows that the full gluon self-energy is the sum of a few terms, namely

\begin{equation}
\Pi_{\rho\sigma}(q; D) = \Pi^q_{\rho\sigma}(q) + \Pi^{gh}_{\rho\sigma}(q) + \Pi_{\rho\sigma}^t(D) +
\Pi^{(1)}_{\rho\sigma}(q; D^2) + \Pi^{(2)}_{\rho\sigma}(q; D^4) + \Pi^{(2')}_{\rho\sigma}(q; D^3),
\label{2.2}
\end{equation}
where $\Pi^q_{\rho\sigma}(q)$ describes the skeleton loop contribution for the quark degrees of freedom
as an analogue of the vacuum polarization tensor in QED.
Note that here and below the superscript or subscript '$q$' means quark (not to be mixed up with the gluon momentum variable $q$). The $\Pi^{gh}_{\rho\sigma}(q)$ describes the skeleton loop contribution associated with the ghost degrees of freedom. Since neither of the skeleton loop integrals depend on the full gluon propagator $D$, they represent the linear contribution to the gluon SD
equation, and $\Pi_{\rho\sigma}^t(D)$ is the so-called constant skeleton tadpole term. $\Pi^{(1)}_{\rho\sigma}(q; D^2)$ represents the skeleton loop contribution,
containing the triple gluon vertices only. Finally, $\Pi^{(2)}_{\rho\sigma}(q; D^4)$ and $\Pi^{(2')}_{\rho\sigma}(q; D^3)$ describe the skeleton two-loop contributions, which combine the triple and quartic gluon vertices. All these quantities are given by the corresponding skeleton loop diagrams in Fig. 1, and they are independent from each other, of course.
The last four terms explicitly contain the full gluon propagators in the corresponding powers symbolically shown above. They form the NL part of the gluon SD equation. The analytical expressions for the corresponding skeleton loop integrals~\cite{25}, in
which the symmetry and combinatorial coefficients and signs have been included, are not important here. We are not going to calculate any of them
explicitly, and thus to introduce into them any truncations/approximations/assumptions or choose some special gauge.

For further purposes it is convenient to present the full gluon self-energy (2.2) as follows:

\begin{equation}
\Pi_{\rho\sigma}(q; D) = \Pi^q_{\rho\sigma}(q) + \Pi^{YM}_{\rho\sigma}(q; D) = \Pi^q_{\rho\sigma}(q) + \Pi^g_{\rho\sigma}(q; D) + \Pi_{\rho\sigma}^t(D),
\end{equation}
where let us remind that the superscript 'YM' stands for the purely Yang-Mills part of QCD. The explicit expression for the tadpole term is

\begin{equation}
\Pi_{\rho\sigma}^t(D) \sim \int d^4 k D_{\alpha\beta}(k) T^0_{\rho\sigma\alpha\beta} = g_{\rho\sigma} \Delta^2_t(D) =
[T_{\rho\sigma} (q) + L_{\rho\sigma}(q)]\Delta^2_t(D),
\end{equation}
where $L_{\rho\sigma}(q) = q_{\rho} q_{\sigma} / q^2$, so that the tadpole term contributes into the both transverse and longitudinal components of the
full gluon propagator via eq.~(2.1). The gluon part denoted as $\Pi^g_{\rho\sigma}(q; D)$ is the sum of all other terms in eq.~(2.2), namely

\begin{equation}
\Pi^g_{\rho\sigma}(q; D) = \Pi^{gh}_{\rho\sigma}(q) + \Pi^{(1)}_{\rho\sigma}(q; D^2) +
\Pi^{(2)}_{\rho\sigma}(q; D^4) + \Pi^{(2')}_{\rho\sigma}(q; D^3).
\end{equation}

All the quantities which contribute to the full gluon self-energy eq.~(2.3), and hence eqs.~(2.4)-(2.5), are tensors, having the dimensions of mass squared. All these skeleton loop integrals are therefore quadratically divergent in the perturbative regime, and so they are assumed to be regularized.
We note, contrary to QED, QCD being a non-abelian gauge theory can suffer from the severe IR singularities in the $q^2 \rightarrow 0$ limit due to the self-interaction of massless gluon modes. Thus, all the possible subtractions at zero may be dangerous \cite{2}. That is why in all the quantities below the dependence on the finite (slightly different from zero) dimensionless subtraction point $\alpha$ is to be understood. In other words, all the subtractions at zero and the Taylor expansions around zero should be understood as the subtractions at $\alpha$ and the structure of the Taylor expansions near $\alpha$, where they are justified to be used. From the technical point of view, however, it is convenient to put formally $\alpha=0$ in all the expressions and derivations below, and to restore the explicit dependence on non-zero $\alpha$ in all the quantities only at the later stage. At the same time, in all the quantities where the dependence on the dimensionless ultraviolet (UV) regulating parameter, $\lambda$ and $\alpha$ is not shown explicitly, nevertheless, it should be also
assumed. For example, $\Pi_{\rho\sigma}(q; D) \equiv \Pi_{\rho\sigma}(q; D, \lambda, \alpha)$ and similarly for all other quantities. This means that all the expressions are regularized (they become finite), and thus a mathematical meaning is assigned to all of them. In this connection, let us underline that the tadpole term (2.4) is quadratically UV divergent constant $\Delta^2_t(D)$, but already regularized one from below and above as well as all the other
such kind of constants which will appear in what follows. Within our approach nothing will depend on how exactly these regulating parameters will be introduced. They will disappear from the theory after the NP renormalization program will be performed. For more detailed description of the general
structure and properties of the SD system of equations see the above-cited references. Let us also remind that the whole gluon momentum range
is $q^2 \in [0, \infty)$ and we are working in Minkowski metric $q^2=q^2_0 - \bf q^2$.

\section{Transversity of the full gluon self-energy}

The first step in the renormalization program of any gauge theory is the removal of the UV quadratic divergences (if any) in order to make
the corresponding theory renormalizable. This can be achieved by introducing suitable subtraction scheme in order to separate them from the PT logarithmic divergences. In this connection it is worth mentioning that a preliminary step in this program, namely to regularize our expressions, has been already done by introducing the corresponding regulating parameters $\lambda$ and $\alpha$ in the previous Section II. In fact, they symbolize that the regularization can be performed by any means, but how exactly is not important, as underlined above.

The basic relation to which the full gluon propagator should satisfy is the corresponding ST identity, namely
\begin{equation}
q_{\mu}q_{\nu} D_{\mu\nu}(q) = - i \xi,
\end{equation}
where $\xi$ is the gauge-fixing parameter.
It is a consequence of the color gauge invariance/symmetry of QCD and, as emphasized above, is an exact constraint on any solution  to QCD.
The ST identity (3.1) implies that the general tensor decomposition of the full gluon propagator in the covariant gauge is as follows:
\begin{equation}
D_{\mu\nu}(q) = - i \left[ T_{\mu\nu}(q) d(q^2) + \xi L_{\mu\nu}(q) \right] {1 \over q^2},
\end{equation}
where the invariant function $d(q^2)= d(q^2; \xi)$ is the corresponding Lorentz structure of the full gluon propagator.
Also, throughout this paper we use the standard definitions of $T_{\mu\nu}(q)$ and $L_{\mu\nu}(q)$, see eq.~(2.4).

If one neglects all the contributions to the full gluon self-energy in eq.~(2.1), i.e., putting formally $d(q^2)=1$ in eq.~(3.2), then
one obtains the free gluon propagator, namely
\begin{equation}
D^0_{\mu\nu}(q) = - i \left[ T_{\mu\nu}(q) + \xi_0 L_{\mu\nu}(q) \right] (1 / q^2),
\end{equation}
where $\xi_0$ is the corresponding gauge-fixing parameter. The general ST identity (3.1) will look like
\begin{equation}
q_{\mu}q_{\nu} D^0_{\mu\nu}(q) = - i \xi_0.
\end{equation}

Contracting now the full gluon SD eq.~(2.1) with $q_{\mu}$ and $q_{\nu}$, and doing some algebra on account of the previous relations (3.1)-(3.4),
one obtains

\begin{equation}
q_{\rho}q_{\sigma} \Pi_{\rho\sigma}(q; D) = - {( \xi_0 - \xi) \over \xi \xi_0 } (q^2)^2.
\end{equation}

From this relation it follows that by itself it cannot remove the quadratic UV divergences from the theory, indeed.  As pointed out above, we will achieve this by formulating the suitable subtraction scheme in which the corresponding transverse conditions can be implemented. But whether they will be satisfied (i.e., equal zero) or not requires much more extra care in dealing with them in QCD.

Let us start the formulation of the subtraction scheme for the full gluon self-energy as follows:
\begin{equation}
\Pi^{(s)}_{\rho\sigma}(q; D)= \Pi_{\rho\sigma}(q; D) - \Pi_{\rho\sigma}(0; D)= \Pi_{\rho\sigma}(q; D) - g_{\rho\sigma}\Delta^2(D), \quad \textrm{which by definition,} \quad \Pi^{(s)}_{\rho\sigma}(0;D) = 0,
\end{equation}
and where
\begin{equation}
\Pi_{\rho\sigma}(0; D) = \Pi^q_{\rho\sigma}(0) + \Pi^g_{\rho\sigma}(0; D) + \Pi^t_{\rho\sigma}(D)
= g_{\rho\sigma} \Delta^2(D) = g_{\rho\sigma} [\Delta^2_q + \Delta^2_g(D) + \Delta^2_t(D)]
\end{equation}
is the sum of the corresponding skeleton loop integrals at $q=0$ (see eqs.~(2.3)-(2.5)), while $\Delta^2_g(D)$ itself is the sum of a few
terms at $q=0$ in eq.~(2.5). All of them are quadratically UV divergent, but already regularized constants.
Let us remind that the subtraction at zero is to be understood in a such way that we subtract at $q^2 = \mu^2$ with $\mu^2 \rightarrow 0$ final limit. It is worth noting that this subtraction is equivalent to add zero to the corresponding identity. For example,
$\Pi_{\rho\sigma}(q; D)= \Pi_{\rho\sigma}(q; D) - \Pi_{\rho\sigma}(0; D) + \Pi_{\rho\sigma}(0; D)$ and denoting
$\Pi^{(s)}_{\rho\sigma}(q; D)= \Pi_{\rho\sigma}(q; D) - \Pi_{\rho\sigma}(0; D)$, one obtains (3.6).
This means that our subtraction scheme change nothing in the initial skeleton loop expressions.

Contracting now (3.6) with $q_\rho$ and $q_\sigma$, one obtains

\begin{equation}
q_{\rho}q_{\sigma} \Pi^{(s)}_{\rho\sigma}(q; D) = - {( \xi_0 - \xi) \over \xi \xi_0 } (q^2)^2 - q^2 \Delta^2(D).
\end{equation}

The general decompositions of the gluon self-energy and its subtracted counterpart into the independent tensor structures look like
\begin{eqnarray}
\Pi_{\rho\sigma}(q) &=& - T_{\rho\sigma}(q) q^2 \Pi_t(q^2) + q_{\rho} q_{\sigma} \Pi_l(q^2), \nonumber\\
\Pi^{(s)}_{\rho\sigma}(q) &=& - T_{\rho\sigma}(q) q^2 \Pi^{(s)}_t(q^2) + q_{\rho} q_{\sigma} \Pi^{(s)}_l(q^2),
\end{eqnarray}
where in all the quantities the dependence on $D$ is omitted, for simplicity, and will be restored below when necessary.
Here and everywhere below all the invariant functions are dimensionless ones of their argument $q^2$: otherwise they remain arbitrary. However,
both invariant functions $\Pi^{(s)}_t(q^2)$ and $\Pi^{(s)}_l(q^2)$ cannot have power-type singularities (or, equivalently, pole-type ones) at small $q^2$, since $\Pi^{(s)}_{\rho\sigma}(0) =0$ by definition in eq.~(3.6).
Thus, one has the two transverse conditions (3.5) and (3.8) for the four invariant functions, which appear in the decompositions (3.9).

Substituting both decompositions into the subtraction (3.6), and doing some algebra on account of the transverse conditions (3.5) and (3.8),
one arrives at
\begin{eqnarray}
\Pi^{(s)}_t (q^2) &=&  \Pi_t(q^2) + {\Delta^2(D) \over q^2}, \nonumber\\
\Pi^{(s)}_l(q^2) &=& \Pi_l(q^2) - {\Delta^2(D) \over q^2} = - {( \xi_0 - \xi) \over \xi \xi_0 } - {\Delta^2(D) \over q^2},
\end{eqnarray}
where the invariant function $-\Pi^{(s)}_t (q^2)$ may still have the logarithmic divergences only in the PT, since all the quadratic UV divergences
summarized into the scale parameter $\Delta^2(D)$, have been already subtracted from the initial invariant function $-\Pi_t(q^2)$.
Then for the full gluon self-energy one gets

\begin{equation}
\Pi_{\rho\sigma}(q) = - T_{\rho\sigma}(q) \left[ q^2 \Pi^{(s)}_t(q^2) - \Delta^2(D) \right] - L_{\rho\sigma}{( \xi_0 - \xi) \over \xi \xi_0 } q^2
\end{equation}
and substituting it into the gluon SD eq.~(2.1), one obtains
\begin{equation}
D_{\mu\nu}(q) = D^0_{\mu\nu}(q) - D^0_{\mu\rho}(q)i T_{\rho\sigma}(q) \left[ q^2 \Pi^{(s)}_t(q^2) -  \Delta^2(D) \right] D_{\sigma\nu}(q)
- D^0_{\mu\rho}(q)i L_{\rho\sigma}(q) {( \xi_0 - \xi) \over \xi \xi_0 } q^2 D_{\sigma\nu}(q),
\end{equation}
which along with the general transverse conditions (3.5) and (3.8) represent the system of equations for the full gluon propagator,
explicitly depending on the mass scale parameter $\Delta^2(D)$.

However, concluding let us underline that contracting it with  $q_{\mu}$ and $q_{\nu}$, one obtains identities $\xi =\xi$ and $\xi_0 = \xi_0$, and
not $\xi$ as a function of $\xi_0$, i.e., $\xi = f(\xi_0)$. In fact, the expression (3.12) is not an equation, but it is an identity! Thus, the transverse relations (3.5) and (3.8) failed to find the function $\xi = f(\xi_0)$ and, at the same time, to change the nature of the expression (3.12).
The important conclusion then is as follows: the only way to get out of these troubles (getting us nowhere) is to satisfy (i.e., put zero) at
least one of them. Only this will
make from the expression (3.12) an equation for the full gluon propagator, and thus to fix the function $\xi = f(\xi_0)$ as well (see below).

\subsection*{Preservation of the exact gauge symmetry and the role of the ghost term}

Let us now show in detail how the $SU(3)$ color gauge symmetry of the QCD Lagrangian might be preserved/saved in its ground state.
By the substitution of the general decompositions (3.2) and (3.3) into eq.~(3.12) and doing some algebra, one obtains

\begin{equation}
D_{\mu\nu}(q) = - i \left[ { 1 \over 1 + \Pi^{(s)}_t(q^2) -  \Delta^2(D) / q^2 } T_{\mu\nu}(q)  + \xi L_{\mu\nu}(q) \right] {1 \over q^2},
\end{equation}
i.e., we express the gluon invariant function $d(q^2)$ in terms of $\Pi^{(s)}_t(q^2)$ and $\Delta^2(D)$ but the gauge-fixing parameter
$\xi$ remains the same. However, if it is not changed, i.e., is not determined/fixed as a function of $\xi_0$, then from the
second of the relations (3.10) it follows that the mass scale parameter
$\Delta^2(D)$ should be disregarded on a general ground, since the invariant function $\Pi^{(s)}_l(q^2)$ cannot have the
pole-types singularities, by definition. This means that the UV divergent, but already regularized constant $\Delta^2(D)$ is to be put
zero everywhere $\Delta^2(D)=0$. Then form (3.7) it follows

\begin{equation}
\Delta^2_q = \Delta^2_g(D) = \Delta^2_t(D) =0,
\end{equation}
i.e., each of these constants should be omitted in the theory, since they are independent from each other (in accordance with the decomposition (2.3)).
This means that the both transverse conditions for the full gluon self-energy (3.5) and its subtracted counterpart (3.8) are
equal to each other, namely

\begin{equation}
q_{\rho}q_{\sigma} \Pi_{\rho\sigma}(q; D) = q_{\rho}q_{\sigma} \Pi^{(s)}_{\rho\sigma}(q; D) = - {( \xi_0 - \xi) \over \xi \xi_0 } (q^2)^2.
\end{equation}
The gluon SD eq.~(3.12) becomes

\begin{equation}
D_{\mu\nu}(q) = D^0_{\mu\nu}(q) - D^0_{\mu\rho}(q)i T_{\rho\sigma}(q) q^2 \Pi^{(s)}_t(q^2) D_{\sigma\nu}(q)
- D^0_{\mu\rho}(q)i L_{\rho\sigma}(q) {( \xi_0 - \xi) \over \xi \xi_0 } q^2 D_{\sigma\nu}(q),
\end{equation}
while its 'solution' looks like

\begin{equation}
D_{\mu\nu}(q) = - i \left[ { 1 \over 1 + \Pi^{(s)}_t(q^2)} T_{\mu\nu}(q)  + \xi L_{\mu\nu}(q) \right] {1 \over q^2}.
\end{equation}
The ST identities (3.1) and (3.4) are respected, of course. It is interesting to note that all the quadratically divergent but regularized constants have to disappear from the theory because the corresponding invariant function cannot have the pole-types singularities, while
the corresponding transverse conditions (3.15) are not yet satisfied.

Therefore the relation (3.16) still remains rather an identity than an equation. As underlined above, to make it equation, the one or both transverse
conditions for the full gluon self-energy and its subtracted counterpart should be satisfied. So that in this case from the relations (3.15), one obtains

\begin{equation}
q_{\rho}q_{\sigma} \Pi_{\rho\sigma}(q; D^{PT}) = q_{\rho}q_{\sigma} \Pi^{(s)}_{\rho\sigma}(q; D^{PT}) =0, \quad \textrm{which requires} \quad \xi =\xi_0,
\end{equation}
and vice versa, i.e., if $\xi =\xi_0$ then the relations (3.15) have to be satisfied.

The relation (3.16) becomes equation, namely

\begin{equation}
D^{PT}_{\mu\nu}(q) = D^0_{\mu\nu}(q) - D^0_{\mu\rho}(q)i T_{\rho\sigma}(q) q^2 \Pi^{(s)}_t(q^2) D^{PT}_{\sigma\nu}(q),
\end{equation}
and from it, one arrives at

\begin{equation}
q_{\mu} q_{\nu}(q)D^{PT}_{\mu\nu}(q) = q_{\mu} q_{\nu}D^0_{\mu\nu}(q)  = - i \xi_0,
\end{equation}
and thus the gauge-fixing parameter of the PT full gluon propagator is fixed as follows: $\xi = f(\xi_0)= \xi_0$, in complete agreement with eq.~(3.18).
Its 'solution' (3.17) now looks like

\begin{equation}
D^{PT}_{\mu\nu}(q) = - i \left[ d^{PT}(q^2) T_{\mu\nu}(q)  + \xi_0 L_{\mu\nu}(q) \right] {1 \over q^2}, \quad  d^{PT}(q^2) = { 1 \over 1 + \Pi^{(s)}_t(q^2)},
\end{equation}
where the invariant function $\Pi^{(s)}_t(q^2)=\Pi^{(s)}_t(q^2; D^{PT})$ is regular function of its argument, free from the quadratic UV divergences, but still may have only the logarithmic ones in the PT $q^2 \rightarrow \infty$ limit. Obviously, this equation describes the propagation of massless gluons, since it is singular on the mass-shell $q^2=0$ only. For the explanation of the notation of the full gluon propagator as $D^{PT}$ in this case see concluding remarks in this subsection.
Let us note in advance that the equality (3.20) takes place only for the regularized massless gluon fields, for their renormalized counterparts these gauge-fixing parameters will be different, of course (see Sections IV-V).

The absence of the mass scale parameters in the system of eqs.~(3.18)-(3.21) can be now attributed to the satisfied transverse conditions
(3.18) as well. Combining it with the second of the decompositions (3.9), from the second of the relations (3.10), one arrives at
$q^2 \Pi^{(s)}_l(q^2) = - \Delta^2(D) = 0$ when $\xi =\xi_0$, which also implies (3.14). It is possible to say that just
the satisfied transverse conditions (3.18) decrease the quadratic UV divergences of the corresponding skeleton loop integrals to a logarithmic ones at large $q^2$. In other words, due to (3.18) all the mass scale parameters, having dimensions of mass squared, shown in (3.14), should be removed/disappeared from the theory, i.e., no explicit presence of such kind of parameters in the PT gluon SD equation and thus in the PT gluon propagator as well.
For how precisely the satisfied transverse relations work in order to remove from the theory the quadratically divergent, but regularized constants, see next subsection.

Let us now make a few remarks concerning the role of the ghost term in the satisfied transverse condition (3.18). From the relation (2.3) and in the absence of the tadpole term, its YM part is reduced to the gluon part, defined in eq.~(2.5). It is well-known that in QCD the quark contribution in the relation (2.3) can be made transverse independently from its YM part, and thus the quark constant $\Delta^2_q$ is to be finally removed from the theory.
Then the satisfied transverse condition (3.18) is reduced to
$q_{\rho} q_{\sigma} \Pi^g_{\rho\sigma}(q; D) = q_{\rho} q_{\sigma} [ \Pi^{gh}_{\rho\sigma}(q) + \Pi^{(1)}_{\rho\sigma}(q; D^2)
+ \Pi^{(2)}_{\rho\sigma}(q; D^4) + \Pi^{(2')}_{\rho\sigma}(q; D^3)] = 0$ and thus the gluon constant $\Delta^2_g(D)$ is to be finally removed from the theory
as well. Let us note in advance that these constants have to be finally omitted in any case, i.e., not only in the PT QCD (see derivations in the next subsection). It is worth reminding now that none of these terms can satisfy the above-displaced transverse condition separately from each other.
In other words, it cannot be re-written down as the sum of the satisfied transverse relations for each term.
 The role of ghost degrees of freedom is to cancel the un-physical (longitudinal) component of the full gluon propagator. Therefore this transverse condition
is important for ghosts to fulfill their role, and thus to maintain the unitary of the $S$-matrix in the PT QCD. Just the Faddeev\,--\,Popov ghost contribution, $\Pi^{gh}_{\rho\sigma}(q)$ makes this transverse relation valid.
For the explicit demonstration of how the ghosts guarantee this transverse condition in lower orders of the PT see, for example~\cite{3,4,5}. This should be true in every order of the PT in agreement with the above-mentioned transverse condition where the skeleton gluon loop diagrams are present.
However, from our analysis above it follows that ghost term makes the transverse conditions (3.18) valid if and only if $\xi=\xi_0$. Note, in the
derivation of the transverse relations (3.18) we did not specify the content of the gluon part of the full gluon self-energy, shown in eq.~(2.5).
Therefore the equality $\xi=\xi_0$ is the first
necessary condition, while the presence of the ghost term there is the second sufficient one for the transverse conditions
(3.18) to be valid. Both conditions are important in the PT QCD, of course. However, in the NP QCD, which is main subject of our investigation in this paper,
the equality $\xi=\xi_0$ will not take the place, i.e, will be violated, while the ghost term will be retained in the full gluon self-energy.
Within our approach to NP QCD they will come into the play in the PT $q^2 \rightarrow \infty$ limit, which will lead to $\xi=\xi_0$, as expected.

The system of eqs.(3.18)-(3.21) is free of all the types of the scale parameters having the dimensions of mass squared, forbidden by the exact
gauge symmetry of the QCD Lagrangian. Therefore, it constitutes that the gauge symmetry of the QCD ground state, reflected by this system,
coincides with the symmetry of its Lagrangian. As it was described in \cite{8}, such coincidence
is analogous to QED where the abelian $U(1)$ gauge symmetry of the Lagrangian is the same as of its ground state. That is why we denoted the full gluon propagator in this system as $D^{PT}$ (suppressing the term $\Delta^2(D) / q^2$ in the PT  $q^2 \rightarrow \infty$ limit in the expression (3.13), one arrives at the expression (3.21)). The essential feature of this phenomenon is the equality $\xi = \xi_0$.
In this connection, let us note that one can start from the expression (3.2). It is nothing else but the general decomposition of the tensor function (the full gluon propagator), depending on the one variable, into the independent tensor structures.
Then it will automatically satisfy to the relation (3.1), called the ST identity and treated the gauge-fixing parameter as a some function of $q^2$, i.e.,
$\xi = f(q^2)$. This has nothing to do with the statement that the ST identity is a consequence of the exact gauge symmetry. It becomes of its consequence, indeed, when the gauge-fixing function is fixed by the relation (3.20), as it has been described in this subsection within the PT methodology.
However, unlike QED,  QCD/YM, being a non-abelian gauge theory with the strong coupling constant, is to be treated beyond the PT, especially its ground state.
So that the general case when $\xi \neq \xi_0$ should be investigated in detail, while keeping the ST identities (3.1) and (3.3) valid, anyway.

\subsection*{Violation of the exact gauge symmetry and the role of the tadpole term}

The distinctive feature of the gluon SD eq.~(3.19) is the absence of any mass squared scale parameters in its dynamical and gauge structures.
None such type of parameters is present in its transverse component, reflecting the corresponding dynamics, nor in its longitudinal
counterpart, reflecting the gauge choice. However, such a scale parameter, namely the tadpole term, is explicitly present in the initial
gluon SD eq.~(2.1) and see Fig. 1 as well. By the virtue of the exact gauge symmetry
all the scale parameters should be disregarded on a general ground. It is important to note that the quark and gluon scale
parameters  $\Delta^2_q$ and $\Delta^2_g(D)$, respectively, are not explicitly present in the full gluon self-energy, but appear as a result of the
corresponding gauge-invariant subtraction scheme. At the same time, the tadpole term $\Delta^2_t(D)$ is explicitly present in the full
gluon-self-energy from the very beginning, making thus the direct contributions into the dynamical and gauge structures of the QCD vacuum. If it has to be removed along with other quadratically divergent but regularized constants due to the gauge invariance, then a natural question arises why is it present in the gluon SD eq.~(2.1) at all? Especially knowing that it makes the YM theory explicitly unrenormalizable! and preventing the ghosts to cancel the longitudinal component of the full gluon propagator and thus making it not transverse.
So one may conclude that the system of eqs.~(3.18)-(3.21) hides the role of the tadpole term in the QCD ground state, making it "invisible".
Nothing would have been changed in the derivation of the above-mentioned system of equations, if the tadpole term in the gluon SD equation (2.1)
were not existed, indeed. In other words, it plays no any dynamical role in the preservation of the exact gauge symmetry in the QCD ground state
(one cannot impose any condition of the cancelation of the quark and gluon scale parameters  $\Delta^2_q$ and $\Delta^2_g(D)$ by $\Delta^2_t(D)$).
Let us emphasize that we are discussing the presence/existence of the tadpole term in the vacuum of QCD from the
dynamical ('physical') point of view, but not its combinatorial (mathematical) meaning.

Just in order to disclose/reveal the true role of the tadpole term in the dynamical and gauge structures of the QCD ground state,
let us investigate the subtraction scheme (3.6) for the full gluon self-energy in more detail.
For this purpose it is instructive to remind the initial eq.~(2.3), which is

\begin{equation}
\Pi_{\rho\sigma}(q; D) = \Pi^q_{\rho\sigma}(q) + \Pi^g_{\rho\sigma}(q; D) + g_{\rho\sigma} \Delta^2_t(D),
\end{equation}
on account of eq.~(2.4), and therefore

\begin{equation}
\Pi_{\rho\sigma}(0; D) = \Pi^q_{\rho\sigma}(0) + \Pi^g_{\rho\sigma}(0; D) + g_{\rho\sigma} \Delta^2_t(D).
\end{equation}

The initial subtraction scheme (3.6) holds, namely

\begin{equation}
\Pi^s_{\rho\sigma}(q; D) = \Pi_{\rho\sigma}(q; D) - g_{\rho\sigma} \Delta^2(D)
\end{equation}
where

\begin{equation}
\Pi^s_{\rho\sigma}(q; D) = [\Pi^q_{\rho\sigma}(q) - \Pi^q_{\rho\sigma}(0)] + [\Pi^g_{\rho\sigma}(q; D) - \Pi^g_{\rho\sigma}(0;D)] = \Pi^{q(s)}_{\rho\sigma}(q)
+ \Pi^{g(s)}_{\rho\sigma}(q; D),
\end{equation}
so that $\Pi^{q(s)}_{\rho\sigma}(0) = \Pi^{g(s)}_{\rho\sigma}(0; D) =0$, by definition, and thus $\Pi^s_{\rho\sigma}(0; D)=0$ as well.

The subtraction (3.24), on account of the relations (3.23), can be expressed as follows:

\begin{equation}
\tilde \Pi^{(s)}_{\rho\sigma}(q; D) = \Pi_{\rho\sigma}(q; D) - g_{\rho\sigma}\Delta^2_t(D)
\end{equation}
because of the relations (3.7), and where

\begin{equation}
\tilde \Pi^{(s)}_{\rho\sigma}(q; D) = [\Pi^{q(s)}_{\rho\sigma}(q) + g_{\rho\sigma} \Delta^2_q] + [\Pi^{g(s)}_{\rho\sigma}(q; D) + g_{\rho\sigma} \Delta^2_g(D)].
\end{equation}

In the initial subtraction (3.6) we did not specify the context of the subtracted gluon self-energy,
while in the both subtracted relations (3.24) and (3.26) we did this and they were present in the relations (3.25) and (3.27), respectively.
From these relations it is explicitly seen that the detailed subtracted parts of the
full gluon self-energy are free from the tadpole term $\Delta^2_t(D)$, as expected.
The difference between these two subtracted gluon self-energies is

\begin{equation}
\tilde \Pi^{(s)}_{\rho\sigma}(q) - \Pi^{(s)}_{\rho\sigma}(q) = g_{\rho\sigma}[\Delta_q^2 + \Delta_g^2(D)].
\end{equation}
From this relation one concludes that contrary to $\Pi^{(s)}_{\rho\sigma}(0)=0$, by definition, the auxiliary subtracted term
$\tilde \Pi^{(s)}_{\rho\sigma}(0) = g_{\rho\sigma}[\Delta_q^2 + \Delta_g^2(D)]$, i.e., it is non-zero constant at this stage, which becomes
zero at final stage only (see below).

Our final goal in this subsection is to find such a gluon SD equation which retain the tadpole term in its structure, but will be free from the quark and gluon constants, since the former one is explicitly present in the initial gluon SD eq.~(2.1), while the latter ones not. On the other hand, we already know that for this purpose one of the transverse conditions should be satisfied, i.e., put zero in order to transform the expression (3.12) into the equation for full gluon propagator. In order to distinguish between the constants $\Delta_q^2$, $\Delta_g^2(D)$ and the tadpole term $\Delta_t^2(D)$,
let us begin with the auxiliary/spurious (or, equivalently, detailed) subtracted gluon self-energy (3.27), since it explicitly depends on the quark and gluon
constants only, so that

\begin{equation}
q_{\rho} q_{\sigma} \tilde \Pi^{(s)}_{\rho\sigma}(q; D) = 0.
\end{equation}
It can be reduced to the two independent transverse conditions due to the sum (3.27) as follows:

\begin{eqnarray}
q_{\rho} q_{\sigma} [\Pi^{q(s)}_{\rho\sigma}(q) + g_{\rho\sigma} \Delta^2_q] &=& 0, \nonumber\\
q_{\rho} q_{\sigma} [\Pi^{g(s)}_{\rho\sigma}(q; D) + g_{\rho\sigma} \Delta^2_g(D)] &=& 0.
\end{eqnarray}
It is worth reminding from the previous subsection that such separation of the satisfied transverse conditions, shown in eqs.~(3.30),
is possible in QCD. The quark contribution can be
made transverse (the first of the relations (3.30)) independently from its gluon part (the second of the relations (3.30)), owing to the fact that the current which flows around the closed skeleton quark loop (see the first skeleton loop diagram in Fig. 1) is conserved from the very beginning. This is in complete analogy with QED, where the current flowing around the closed skeleton electron loop (vacuum polarization tensor, which only one contributes into the full photon self-energy) is conserved.

Substituting further the corresponding independent tensor decompositions

\begin{eqnarray}
\Pi^{q(s)}_{\rho\sigma}(q) &=& - T_{\rho\sigma}(q) q^2 \Pi^{q(s)}_t(q^2) + q_{\rho} q_{\sigma} \Pi^{q(s)}_l(q^2), \nonumber\\
\Pi^{g(s)}_{\rho\sigma}(q; D) &=& - T_{\rho\sigma}(q) q^2 \Pi^{g(s)}_t(q^2;D) + q_{\rho} q_{\sigma} \Pi^{g(s)}_l(q^2; D)
\end{eqnarray}
into the satisfied transverse relations (3.30), and doing some algebra, one obtains

\begin{equation}
\Pi^{q(s)}_l(q^2) = - { \Delta^2_q \over q^2}, \quad \quad  \Pi^{g(s)}_l(q^2; D)= - { \Delta^2_g(D) \over q^2}.
\end{equation}
However, both relations are not possible, since $\Pi^{q(s)}_l(q^2)$ and  $\Pi^{g(s)}_l(q^2;D)$
by themselves cannot have power-type singularities at small $q^2$, because of the relations $\Pi^{q(s)}_{\rho\sigma}(0) = \Pi^{g(s)}_{\rho\sigma}(0; D)=0$,
displayed above.
This means that the quadratically UV divergent, but already regularized constant $\Delta^2_q$ and $\Delta^2_g(D)$ are to be disregarded on a general ground, i.e. put zero everywhere

\begin{equation}
\Delta^2_q = \Delta^2_g(D)=0, \quad \textrm{which means that} \quad \Delta^2 (D) = \Delta^2_t(D) \neq 0,
\end{equation}
as it comes out from the relation (3.7), while the tadpole contribution $\Delta^2_t(D)$ remains intact. From the relations (3.32)
and (3.33) one concludes that $\Pi^{q(s)}_l(q^2) = \Pi^{g(s)}_l(q^2; D)= 0$ as well, which means the satisfied transverse relations
$q_{\rho} q_{\sigma}\Pi^{q(s)}_{\rho\sigma}(q) = q_{\rho} q_{\sigma} \Pi^{g(s)}_{\rho\sigma}(q; D) = 0$ take place due to the relations (3.31).
Then from the relations (3.25) it follows that the initial $\Pi^{(s)}_{\rho\sigma}(q)$, defined in eq.~(3.6), is also transverse, i.e.,

\begin{equation}
q_{\rho} q_{\sigma} \Pi^{(s)}_{\rho\sigma}(q; D) = 0.
\end{equation}
From it, and due to the relations (3.9), (3.10) and (3.33), one finally obtains

\begin{eqnarray}
\Pi^{(s)}_t (q^2) &=&  \Pi_t(q^2) + {\Delta_t^2(D) \over q^2}, \nonumber\\
\Pi^{(s)}_l(q^2) &=& - {( \xi_0 - \xi) \over \xi \xi_0 } - {\Delta_t^2(D) \over q^2} = 0.
\end{eqnarray}

In connection with the second of the relations (3.35), let us remind that it can be satisfied if $\xi = \xi_0$, then $\Delta^2_t(D)=0$ as well, since the invariant function $\Pi^{(s)}_l(q^2)$ cannot have the pole-type singularities, by definition. Just these conditions have been used in order to get to the system of eqs.~(3.18)-(3.21), which preserves the exact gauge symmetry of the QCD Lagrangian in its ground state. However, it has another
(more general in our opinion)
solution. It shows the gauge change ($\xi$ is different from $\xi_0$), so that the gauge-fixing parameter for the full gluon propagator $\xi$ becomes the corresponding function of $q^2$, i.e., $\xi= \xi(q^2)$. From the second equation in the relations (3.35), one gets

\begin{equation}
{( \xi_0 - \xi) \over \xi \xi_0 } = - {\Delta_t^2(D) \over q^2},
\end{equation}
precisely which solution will determine the above-mentioned function $\xi= \xi(q^2)$. Its explicit solution
is convenient to postpone until the next section. Using it in the transverse condition (3.5), one arrives at the two independent transverse conditions as follows:

\begin{eqnarray}
q_{\rho}q_{\sigma} \Pi_{\rho\sigma}(q; D) &= &  q^2 \Delta^2_t(D)  \neq 0,  \nonumber\\
q_{\rho}q_{\sigma} \Pi^{(s)}_{\rho\sigma}(q; D) &=& 0,
\end{eqnarray}
instead of the transverse conditions (3.18). That is why we call such system of the transverse conditions as a result of the splintering procedure, since by formally neglecting the tadpole term they become the same, i.e., reduced to the relations (3.18), and thus there is no need in their separate treatment.

The initial gluon SD eq.~(2.1), via the auxiliary eq.~(3.12), now becomes

\begin{equation}
D_{\mu\nu}(q) = D^0_{\mu\nu}(q) - D^0_{\mu\rho}(q)i T_{\rho\sigma}(q) \left[ q^2 \Pi^{(s)}_t(q^2) -  \Delta^2_t(D) \right] D_{\sigma\nu}(q)
+ D^0_{\mu\rho}(q)i L_{\rho\sigma}(q) \Delta^2_t(D) D_{\sigma\nu}(q),
\end{equation}
on account of the relations (3.33) and (3.36). The corresponding ST identities therefore are not equal to each other

\begin{equation}
q_{\mu} q_{\nu}(q)D_{\mu\nu}(q) \neq q_{\mu} q_{\nu}D^0_{\mu\nu}(q), \quad \textrm{which implies} \quad \xi \neq \xi_0,
\end{equation}
and vice versa, i.e., when $\xi \neq \xi_0$ then an equality is impossible.

If the tadpole term is formally omitted in the system of eqs.~(3.37)-(3.39), then it will be reduced to the system of eqs.~(3.18)-(3.21), as pointed
out just above.
So that the system of eqs.~(3.18)-(3.21), preserving the gauge symmetry of the QCD Lagrangian in its ground state, is a particular case
of the general system of eqs.~(3.37)-(3.39), which reflects the violation of the Lagrangian gauge symmetry in its ground state. It is important to
understand that the system of eqs.(3.18)-(3.21) describes a hypothetical situation in QCD, while the system of eqs.~(3.37)-(3.39) describes the real
dynamical and gauge structures of its ground state. However, the solutions of the system of eqs.~(3.18)-(3.21) can be part of the solutions of the system of eqs.~(3.37)-(3.39). For example, in the PT $q^2\rightarrow \infty$ limit, when the contribution $\Delta_t(D) / q^2$ can be neglected (see Section VI).

The final system of eqs.(3.37)-(3.39) will not depend on how precisely one introduces the subtraction scheme.
For example, one can define the subtracted gluon self-energy, instead of (3.27), as follows: $\bar \Pi^{(s)}_{\rho\sigma}(q; D)= \Pi^q_{\rho\sigma}(q) + \Pi^g_{\rho\sigma}(q; D)$ in the initial expression (3.22). Note, this definition also does not  depend on the tadpole term, as requested, and $\bar \Pi^{(s)}_{\rho\sigma}(0; D) =\delta_{\rho\sigma}[\Delta_q^2 + \Delta_g^2(D)]$ as well. But from eq.~(3.27) it follows that $\bar \Pi^{(s)}_{\rho\sigma}(q; D)= \tilde \Pi^{(s)}_{\rho\sigma}(q; D)$. Repeating the derivations, it is easy to show that
$q_{\rho}q_{\sigma} \bar \Pi^{(s)}_{\rho\sigma}(q; D) = q_{\rho}q_{\sigma} \tilde \Pi^{(s)}_{\rho\sigma}(q; D) = 0$.
The uniqueness of our approach (in order to establish
the true dynamical and gauge structures of the QCD ground state) will be demonstrated from the general point of view in Appendix A.
The uxiliary/spurious scheme (3.26) described above makes it possible to distinguish between the quark and gluon constants, which are not present in the initial gluon SD eq.~(2.1), and the tadpole constant, which is explicitly present in it. The initial subtraction scheme (3.6) failed to do this. At the same time,
the detailed subtraction scheme finally leads to the satisfied transverse relation (3.34) for the initially subtracted gluon self-energy (3.24), as expected.

Concluding, it is important to emphasize once more that we are under an obligation to satisfy the transverse relation for the subtracted gluon self-energy
(defined in any possible way). Reminding that this is necessary to do in order to make from the expression (3.12) an equation for the full gluon propagator.
In other words, none of the above-mentioned satisfied transverse relations have been introduced by hand, but they have been implemented
into the our formalism in a self-consistent way. The fact of the matter is that the system of the transverse relations (3.37) has been derived.

\subsection*{Preliminary remarks}

The formalism developed in the previous subsection clearly shows that the role of the tadpole term $\Delta^2_t(D)$ in the dynamical and
gauge structures of the QCD ground state is different from those of the quark $\Delta^2_q$ and gluon $\Delta^2_g(D)$ constants.
Contrary to the gauge symmetry preservation, investigated in the first subsection of Section III, the treatment
of the tadpole term cannot be put on the same footing as the quark and gluon constants, which should be omitted in the theory anyway.
Let us remind that the removal of the quark and gluon constants are due to the properties of the corresponding invariant functions, see relations (3.32),
while such an invariant function does not exist for the tadpole term itself. The derived splintering transverse relations (3.37) make it possible
for the tadpole term  $\Delta^2_t(D)$ to remain in the gluon SD eq.~(3.38) and thus it will appear in the full gluon propagator as well.
The gluon SD eq.~(3.38) is equivalent to the initial gluon SD eq.~(2.1), while eq.~(3.19) is not. In other words, eq.~(3.19)
drastically distorts the true dynamical and gauge structures of eq.~(2.1), while eq.~(3.38) preserves it.
The only difference is that the skeleton loop contributions to eq.~(2.1) are taken into account in terms of the corresponding
invariant function and the tadpole term in the full gluon SD eq.~(3.38). It is much more convenient for its solution and developing
the corresponding NP MP renormalization program than the initial eq.~(2.1).
There is no doubt that the exact gauge symmetry of the QCD Lagrangian is dynamically broken in its ground state by the explicit presence
of the tadpole term in the full gluon self-energy. Precisely the system of eqs.~(3.37)-(3.39) reflects this effect.
To reveal the tadpole term's true role in full details was a clue to this discovery. Previously it has been described in some details in~\cite{8}
(and see our references therein as well).
The existence of the tadpole term is a bright signal/evidence that the real dynamical and gauge structures of the QCD ground state are not
so simple as it is required by the exact gauge symmetry of its Lagrangian.

Let us present a few important observations supporting our general statements made just above.

\begin{itemize}
\item[(A).] Any deviation of the full gluon propagator from the free one requires the presence of the mass squared scale parameter
on the general dimensional ground. Even in the AF regime there is a scale violation
\begin{equation}
D_{\mu\nu}(q) \sim g_{\mu\nu} \Bigl[ { g^2 \over  1 + g^2 b_0 \ln(q^2/ \Lambda^2_{QCD}) } \Bigr] (1 / q^2),
\end{equation}
where  $g^2$ is the coupling constant and $b_0$ is the color group factor. This expression presents the summation of the so-called main PT logarithms
in QCD and written down in the 't Hooft\,--\,Feynman gauge \cite{3,4,5,8}.

\item[(B).] However, the mass scale parameter $\Lambda^2_{QCD}$, which determines the non-trivial PT dynamics in the QCD vacuum, cannot be generated
by the PT itself. Due to the renormalization group equations arguments \cite{2}, any mass to which can be assigned some physical sense disappears
according to
\begin{equation}
M  \sim \mu \exp(- 1 /  b_0 g^2), \quad \quad g^2 \rightarrow 0,
\end{equation}
where $\mu$ is the arbitrary renormalization point. In other words, in every order of the PT it vanishes. So that it has to come from the IR region,
since non a finite mass can survive in the PT $q^2 \rightarrow \infty$ limit. Such kind of mass can be of the NP dynamical origin only.

\item[(C).] Due to the self-interaction of multiple massless gluon modes, QCD suffers from dangerous IR singularities (more severe than $(1/q^2)$ PT one).
The existence of the severe IR singularities also requires the presence of the mass scale parameter $\Delta^2$ in the full gluon propagator
(but "dressed" gluon itself remains massless), for example
\begin{equation}
D_{\mu\nu}(q) \sim T_{\mu\nu} {1 \over q^2} \sum_{k=0}^{\infty} \Bigl({ \Delta^2 \over q^2 } \Bigr)^{k+1} \Phi_k,
\end{equation}
where the arbitrary coefficients $\Phi_k$ by themselves are the sums of the infinite number of terms~\cite{49}. This expression presents the summation
of all the possible severe IR singularities which can be taken into account by the full gluon propagator. It is worth noting that the mass scale
parameter $\Delta^2$ may contribute into the longitudinal component as well. How to deal with such severe IR singularities for the first
time has been described in~\cite{8} and in detail in our recent work~\cite{59} (see our references therein as well).

\item[(D).] In general, the symmetries of the Lagrangian of the quantum/classical field gauge theory may not coincide with the symmetries of its ground state (vacuum).

\item[(E).] The color charge is not conserved in QCD. The derived splintering expressions (3.37) explicitly reflects this fact analytically,
while respecting the corresponding ST identities for the full gluon propagator and its free counterpart.
\end{itemize}

The invariant function $\Pi^{(s)}_t(q^2)$ associated with the transverse component of the full gluon SD eq.~(3.38), and thus with the transverse component
in the full gluon propagator, is regular at zero and only logarithmical divergent at $q^2 \rightarrow \infty$. So it can be subject of the PT renormalization program (and it is not our problem here).
However, the tadpole term which contributes into the both components of the gluon SD eq.~(3.38), and thus contributes into the transverse and
the longitudinal component as well of the full gluon propagator, is quadratically divergent, but regularized constant. It cannot be renormalized
by the PT technics. Therefore the NP renormalization program has to be developed (just it is our problem here). For this purpose
it is useful to introduce the notation as follows:
\begin{equation}
 \Delta^2_t(D) = M^2,
\end{equation}
since it has the dimension of mass squared at any $D$, and therefore we denote it as $M^2$. Let us remind that it is quadratically UV divergent constant,
but regularized one. Its renormalized version will be called the mass gap since it will separate massive solution for the full gluon propagator from its massless counterpart in what follows. First of all, the above-mentioned NP renormalization program should be performed for $M^2$ in order to make the theory renormalizable, but before the formulation of the mass gap approach to QCD should be completed.

Concluding, a few more remarks are in order. From our analysis one can decide that not losing generality we can omit the quark degrees of freedom
below and investigate only the purely YM part of QCD. We have discussed in detail some important aspects of the color gauge structure
of the gluon SD equation in the YM gauge theory, but without any use of the PT. In obtaining these results no specific regularization schemes (preserving or not gauge invariance) has been used. No special gauge choice and no any truncations/approximations/assumptions  have been made either. Only
analytic derivations have been done, such as the decomposition into the independent tensor structures, the subtractions, etc. We have shown in the most general and unique way that the gauge symmetries of the QCD Lagrangian and its ground state are not the same, indeed.

\section{The mass gap approach to QCD}

In order to calculate the physical observables in QCD from first principles, we need the full gluon propagator rather than the full gluon self-energy.
As emphasized earlier, the basic relations to which the full gluon propagator and its free counterpart should satisfy are the corresponding ST identities (3.1) and (3.4), respectively. They are consequence of the color gauge invariance/symmetry of QCD. However, by themselves they cannot remove the UV divergences
from the theory, as it has been shown above.
We have achieved this by formulating the suitable subtraction scheme in which the corresponding transverse conditions have been implemented.
If some equations, relations or the regularization schemes, etc.  do not satisfy them automatically, i.e., without any additional conditions, then they should be modified and not the identity (3.1). In other words, all the relations, equations, regularization schemes, etc. should be adjusted to it and not vice versa. For example, the transverse condition for the full gluon self-energy (3.37) is violated, but, nevertheless,
the general form of the ST identity (3.1) is to be maintained despite the massless or massive gluon fields are considered.
Saving the mass scale parameter in the transverse part of the full gluon propagator necessary leads to the gauge-changing
phenomenon, i.e., makes it possible to fix the function $\xi= f(\xi_0)$.
It is the legitimate procedure since the full gluon propagator is defined up to its longitudinal part only. This is true for its equation of motion as well. As explained above, the mass scale parameter is very much needed in the transverse part of the full gluon propagator in order to correctly reflect the true dynamical structure of the QCD ground state.

Let us begin with the gluon SD eq.~(3.38), which can be equivalently re-written down as follows:

\begin{equation}
D_{\mu\nu}(q) = D^0_{\mu\nu}(q) - D^0_{\mu\rho}(q)i T_{\rho\sigma}(q) \left[ q^2 \Pi(q^2; D) -  M^2 \right] D_{\sigma\nu}(q)
+ D^0_{\mu\rho}(q)i L_{\rho\sigma}(q) M^2 D_{\sigma\nu}(q),
\end{equation}
where $\Pi(q^2; D)$ is a replacement (for convenience) of the initial $\Pi^{(s)}_t(q^2; D)$ in eq.~(3.38). It is worth remanding that it is regular at zero
and may have only logarithmic divergences in the PT $q^2 \rightarrow \infty$ limit.
Combining this equation with the decompositions (3.2) and (3.3), one obtains
\begin{equation}
d(q^2) = {1 \over 1 + \Pi(q^2; D) - (M^2 / q^2)}.
\end{equation}
This relation is not a solution for the full gluon invariant function,
but rather some kind of the transcendental equation for different invariant functions $d(q^2), \ \Pi(q^2; D)$ and the mass gap $M^2$,
i.e., $d=f(D(d))$. Nevertheless, from this expression is clearly seen that in the PT $q^2 \rightarrow \infty$ regime, the mass gap term contribution
$(M^2 / q^2)$ can be neglected, but the invariant function $\Pi(q^2; D)$ may still depend on $M^2$ under the PT logarithms.
At the same time, in the NP region of finite and small gluon momenta this term is dominant, and the dependence of $d(q^2)$ on $M^2$ in this case may be much more complicated due to the transcendental character of eq.~(4.2). That is why it should be kept 'alive' on a general ground, indeed.
However, as pointed out above, keeping it 'alive' makes the YM theory unrenormalizable. So the problem arises how to make the theory renormalizable
in this case as well (see next sections).

It is instructive now to show explicitly the gauge-changing phenomenon, mentioned above.
Contracting the full gluon SD eq.~(4.1) with $q_{\mu}$ and $q_{\nu}$, and substituting its result into the general ST
identity (3.1), one arrives at
\begin{equation}
q_{\mu}q_{\nu} D_{\mu\nu}(q) = - i \xi_0 \left( 1 + \xi  {M^2 \over q^2} \right) = - i \xi,
\end{equation}
which solution is
\begin{equation}
\xi \equiv \xi(q^2) = { \xi_0 q^2 \over q^2 - \xi_0 M^2},
\end{equation}
i.e., in this case the gauge-fixing parameter becomes the function $\xi \equiv \xi(q^2)$ and not a constant like $\xi_0$.
This is in fair agreement with our discussion just above. Let us point out that the expression (4.4) satisfies the gauge changing
condition (3.36), which is as it should be.
From (4.4) it is clearly seen that in the PT $q^2 \rightarrow \infty$ regime, which effectively equivalent to the formal $M^2=0$ limit and vice versa,
the gauge-fixing parameter $\xi$ goes to $\xi_0$, as expected.
Thus, behind the general inequality $\xi \neq \xi_0$ is the tadpole term as its dynamical source, when its contribution can be neglected
only then $\xi = \xi_0$. In this connection, let us note that in the previous and this sections the equality $\xi = \xi_0$ holds
only for the regularized gluon fields. For the renormalized full gluon propagator and the free one their gauge-fixing parameters will be different
even in the formal $M^2=0$ limit, see next section.

Substituting (4.2) and (4.4) into the general decomposition (3.2) for the full gluon propagator, one finally obtains
\begin{equation}
D_{\mu\nu}(q) =  {- i T_{\mu\nu}(q) \over q^2 + q^2 \Pi(q^2; D) - M^2 }  - i L_{\mu\nu}(q) { \lambda^{-1} \over q^2 - \lambda^{-1} M^2},
\end{equation}
where we have introduced the useful notation, namely $\xi_0 = \lambda^{-1}$~\cite{4} (not to be confused with the dimensionless UV regulating
parameter, mentioned in Section II). The corresponding ST identity now becomes
\begin{equation}
q_{\mu}q_{\nu} D_{\mu\nu}(q) = - i \xi(q^2)  = - i { \lambda^{-1} q^2 \over (q^2 - \lambda^{-1} M^2))}.
\end{equation}

In what follows we call it as the generalized gauge since it depends on the tadpole term $M^2$, and when it is zero, one recovers the gauge-fixing parameter for the free massless gluon propagator. To our best knowledge the generalized gauge (4.6) has been derived for the first time in a such new manner.
In the generalized gauge the gauge-fixing parameter $\lambda^{-1}$ can vary continuously from zero to infinity.
The functional dependence of the generalized gauge-fixing parameter $\xi$ (4.4) is fixed up to an arbitrary gauge-fixing parameter $\xi_0 = \lambda^{-1}$. Unless we fix it, and thus $\xi$ itself, we will call such situation as the general gauge dependence (GGD), see (3.1), (4.5), and (4.6).
Choosing $\xi_0 = \lambda^{-1}$ explicitly, we will call such situation as the explicit gauge dependence (EGD).
For example, $\xi_0 = \lambda^{-1}=0$ is called the unitary (Landau) gauge, $\xi_0 = \lambda^{-1}=1$ is called the t' Hooft--Feynman gauge, etc.~\cite{26,27,28,29,30}. The formal $\xi_0 = \lambda^{-1}= \infty$ limit is called as the canonical gauge in~\cite{29}, and its properties
will be discussed in detail below. This distinction seems a mere convention, but, nevertheless, it is useful one in QCD because
of the presence of the mass scale parameter in its ground state.
The generalized gauge directly follows from the GGD/EGD formalism within the mass gap approach to QCD. It requires that there is no other functional expression for $\xi$, apart from given by the relation (4.4) at finite  $\xi_0 = \lambda^{-1}$, in the full gluon propagator (4.5)
and the corresponding ST identity (4.6) for the gluon fields in this case. In other words, unlike the expression (4.2), which is the
NL transcendental relation, the expression (4.4) is the known function, which determines  $\xi = \xi(q^2)$.
The system of equations, consisting of the gluon
SD eq.~(4.1), the expression (4.5) and the ST identity (4.6), constitutes that the $SU(3)$ color gauge symmetry
of the QCD Lagrangian is not a symmetry of its ground state. It is important to note that the tadpole term enters the full gluon self-energy linearly,
see, for example, the expression (2.3). However, in the full gluon propagator (4.5) it appears in the completely different NL way. This happened
because its contribution has been summed up with the help of the gluon SD eq.~(4.1). Let us underline once more that
the tadpole term contributes into the transcendental expression, associated with the transverse component of the full gluon propagator (4.5).
At the same time, its contribution into the longitudinal component of eq.~(4.5) renders it to the known function of $q^2$ and $M^2$.
That is why there is no equivalence between the PT $q^2 \rightarrow \infty$ and formal $M^2=0$ limits in the former case, while this equivalence
persists in the latter one. Eq.~(4.5) shows the general structure of the full gluon propagator with the mass gap or without mass gap if it is formally put
zero. In what follow we put $\Pi(q^2; D)= \Pi(q^2)$, for convenience.

\section{NP renormalization of the massive gluon propagator}

Let us now perform the renormalization program for the full gluon propagator, presented in the relations (4.5) and (4.6).
If the denominator in eq.~(4.5) may have a pole at the point $q^2=M_g^2$, then one obtains
\begin{equation}
M^2 =[ 1  +  \Pi(M^2_g)]M^2_g = {\tilde Z}_3^{-1} M^2_g,
\end{equation}
where
\begin{equation}
{\tilde Z}_3^{-1} \equiv {\tilde Z}_3^{-1}(M^2_g)= [ 1  +  \Pi(M^2_g)],
\end{equation}
and it can be treated as the mass gap NP multiplicative (MP) renormalization constant, while $M^2$ being the 'bare' gluon mass, and $M^2_g$
being its renormalized counterpart, i.e., the mass gap itself, as pointed out above. It can be interpreted as a gluon true pole mass, since
it is exactly defined by the relation (5.1). Due to its transcendental character (since it enters the transcendental eq.~(4.2)), the function $\Pi(q^2)$
at the point $q^2=M_g^2$ may be indeed quadratically divergent, when the UV regulating parameter goes to infinity.

Let us further expand the invariant function $\Pi(q^2)$ around the pole position, namely
\begin{equation}
\Pi(q^2) = \Pi(M^2_g) + (q^2 - M^2_g) \Pi'(M^2_g) + (q^2 - M^2_g)^2 \Pi''(M^2_g) + ...
\end{equation}
Evidently, this expansion is the corresponding Taylor series. Substituting all these expressions back into the denominator
and doing some algebra, one obtains
\begin{equation}
q^2 + q^2 \Pi(q^2) - M^2 = (q^2 - M^2_g) Z_3^{-1} [ 1 + Z_3 {\tilde \Pi}(q^2)],
\end{equation}
where
\begin{equation}
Z_3^{-1} \equiv Z_3^{-1}(M^2_g) = 1 + \Pi(M^2_g) + M^2_g \Pi'(M^2_g) = {\tilde Z}_3^{-1} + M^2_g \Pi'(M^2_g),
\end{equation}
and it can be treated as the gluon wave function NP MP renormalization constant. The invariant function, ${\tilde \Pi}(q^2)$ is defined as follows:
\begin{equation}
{\tilde \Pi}(q^2) = (q^2 - M^2_g) \left[ \Pi'(M^2_g) + q^2 \Pi''(M^2_g) + q^2(q^2 - M^2_g) \Pi'''(M^2_g) + ... \right].
\end{equation}
In the derivation of the relations (5.4)-(5.6) we have used the following obvious identity $q^2 = (M_g^2 + q^2 - M^2_g)$ in the term
with the first derivative $\Pi'(M^2_g)$. Let us point out that the invariant function, ${\tilde \Pi}(q^2)$ in eq.~(5.6) has the following
interesting properties as
\begin{equation}
{\tilde \Pi}(M^2_g)=0, \quad \quad {\tilde \Pi}(0) = - M^2_g \Pi'(M^2_g),
\end{equation}
so that it vanishes at the pole position, i.e., on the mass-shell $q^2=M^2_g$. Let us note that the same effect will be achieved, i.e., ${\tilde \Pi}(q^2)=0$, if one formally puts all the derivatives $\Pi^{(n)}(M^2_g)=0, \ n=1,2,3...$, as it follows from the expansion (5.6). This means
that the difference between the NP MP renormalization constants $Z_3^{-1}$ and ${\tilde Z}_3^{-1}$ can be effectively neglected on the mass-shell
up to leading order in powers of small $\Pi^{(1)}(M^2_g) = \Pi'(M^2_g)$ or finite but small $M^2_g$. From the definition (5.5) it follows
$Z_3 {\tilde Z}_3^{-1} = 1 - Z_3  M^2_g \Pi'(M^2_g) = 1 + \sum_{n=1}^{\infty} (-1)^n[M_g^2 \Pi'(M^2_g){\tilde Z}_3]^n$, indeed.

Substituting all these relations into eqs.~(4.5)-(4.6), one finally arrives at the following system for the
renormalized massive gluon propagator $D^R_{\mu\nu}(q)$ within the GGD formalism, namely
\begin{equation}
D^R_{\mu\nu}(q) = { - i Z_3  {\tilde Z}^{-1}_3 \over (q^2 - M^2_g)[ 1 + Z_3 {\tilde \Pi}(q^2)]} T_{\mu\nu}(q)
- i L_{\mu\nu}(q)  {{\tilde\lambda}^{-1} \over [q^2 - {\tilde\lambda}^{-1} M_g^2]},
\end{equation}
where $D^R_{\mu\nu}(q)$ is defined as follows:
\begin{equation}
D^R_{\mu\nu}(q) = {\tilde Z}^{-1}_3 D_{\mu\nu}(q), \quad  \tilde\lambda = \lambda {\tilde Z}_3,
\quad Z_3 {\tilde Z}_3^{-1} = 1 - Z_3  M^2_g \Pi'(M^2_g),
\end{equation}
and the massive gluon propagator (5.8) is shown in Fig. 2. The renormalized ST identity becomes
\begin{figure}[h]
\begin{center}
\includegraphics[width=15.0truecm]{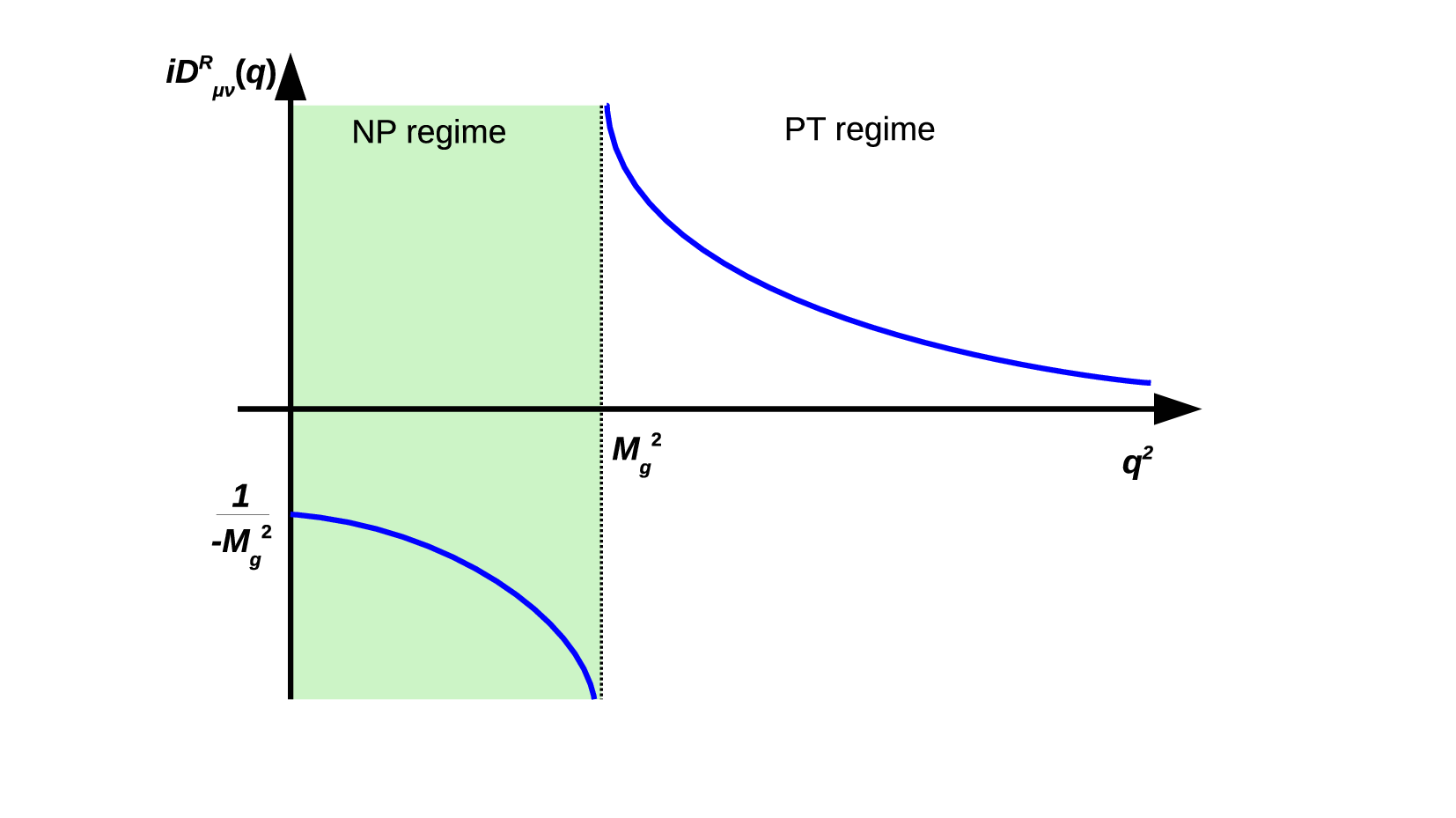}
\caption{The massive gluon propagator (5.8) at finite $\tilde{\lambda}^{-1}$ as a function of $q^2$.
Note, the values at $q^2=0$ and at $q^2=M^2_g$ are exactly and gauge-independently defined.}
\label{fig:2}
\end{center}
\end{figure}

\begin{equation}
q_{\mu}q_{\nu} D^R_{\mu\nu}(q)  = - i \xi^R(q^2) = - i {{\tilde\lambda}^{-1} q^2 \over q^2 -  {\tilde\lambda}^{-1} M_g^2}.
\end{equation}
From now on all the quadratically UV divergences disappear from the theory. Along with the arbitrary subtraction point they have been incorporated
into the NP MP renormalization
constants (5.2) and (5.5). The invariant function ${\tilde \Pi}(q^2)$ and the full gluon propagator itself are regular at zero. Moreover, in the most
important case, i.e., on the mass-shell $q^2=M^2_g$, the invariant function is zero, see eq.~(5.7). Thus, one can conclude
that the YM theory is NP renormalizable within our approach since all the quantities in the full gluon propagator (5.8), and hence in
the ST identity (5.10), are expressed in terms of the renormalized quantities only. The dependence on the NP MP renormalization constants
should disappear in QCD through the corresponding identity, which includes quark degrees of freedom. This is beyond the scope of the present investigation
(see discussion in Section IX as well).

Neglecting the contribution from the regular part of the full gluon-self energy, i.e., putting formally
\begin{equation}
\Pi(q^2) = {\tilde \Pi}(q^2) = 0, \ Z_3 = {\tilde Z}_3 = 1, \ {\tilde\lambda} = \lambda, \ M^2_g = M^2,
\end{equation}
then from eq.~(5.8) for the free massive vector boson propagator in the generalized gauge, one obtains
\begin{equation}
D^0_{\mu\nu}(q; M^2_g) =  { - i \over (q^2 - M^2_g)} \left[ g_{\mu\nu} - (1 - \lambda^{-1})
{q_{\mu}q_{\nu} \over (q^2 - \lambda^{-1} M^2_g)} \right],
\end{equation}
which satisfies the same ST identity (5.10) as a function of $\lambda^{-1}$, as expected. Evidently, in the $M^2_g=0$ limit it becomes the free massless gluon propagator (3.3). This free massive vector particle propagator for the first time has been used in the investigations of the different models with spontaneously broken gauge theories in~\cite{4,27,28,29,30}. In~\cite{4} the last expression has been called as one being written down in the Stueckelberg gauge.

Within the EGD formalism let us show the free massive vector boson propagator (5.12) in some different finite gauges, such as
\begin{equation}
\textrm{unitary \ (Landau) \ gauge,} \ \  \lambda^{-1}=0: \quad  D^0_{\mu\nu}(q; M^2_g) = { - i \over (q^2 - M^2_g)}T_{\mu\nu}(q),
\end{equation}
\begin{equation}
\textrm{t'Hooft\,--\,Feynman \ gauge,} \ \ \lambda^{-1}=1:  \quad  D^0_{\mu\nu}(q; M^2_g) = { - i \over (q^2 - M^2_g)}g_{\mu\nu}.
\end{equation}

\section{Asymptotics of the massive full gluon propagator}

Let us now investigate the structure of the full gluon propagator with the gluon pole mass (5.8) in the $q^2 \rightarrow 0$ limit.
Then one arrives at
\begin{equation}
D^R_{\mu\nu}(q) = {  i {\tilde Z}^{-1}_3 \over M^2_g \left[ Z_3^{-1} + {\tilde \Pi}(0) \right]} T_{\mu\nu}(q) + i L_{\mu\nu} (q) { 1 \over M_g^2},
\quad q^2 \rightarrow 0
\end{equation}
Taking now in account relations (5.5) and (5.7), one finally obtains
\begin{equation}
iD^R_{\mu\nu}(0) = - { 1 \over M^2_g} T_{\mu\nu}(q) - L_{\mu\nu} (q) { 1 \over M_g^2} =  - { 1 \over M^2_g} g_{\mu\nu},
\end{equation}
and this result is exact, gauge-independent and regular at $q^2 = 0$ as Fig. 2 presents. Evidently, this value coincides with the expression (5.12)
at any gauge in the $q^2 \rightarrow 0$ limit, as it needs be.

It is worth now to explicitly show that the system of eqs.~(3.18)-(3.21) follows from the system of eqs.~(5.8)-(5.10) at $M_g^2=0$.
Obviously, the mass-shell and the pole position in the massless case is defined at $q^2=0$.
Let us remind that this limit means that we are putting the treatment of the tadpole term on the same footing as the quark and gluon regularized constants, i.e., all the quadratically divergent but regularized scale parameters are to be disregarded and thus removed from the theory, see eq.~(3.14).
In other words, the exact gauge symmetry of the QCD Lagrangian is to be restored in its ground state in this case.
So letting $M_g^2=0$ at finite ${\tilde\lambda}^{-1}$ in eqs.~(5.8)-(5.10), one obtains
\begin{equation}
D^R_{\mu\nu}(q) = { - i \over q^2 \left[ 1 + Z_3 (0)\Pi^r(q^2)\right] } T_{\mu\nu}(q)
- i {\tilde\lambda}^{-1} L_{\mu\nu} (q) { 1 \over q^2},
\end{equation}
where the invariant function $\Pi^r(q^2) = {\tilde \Pi}(q^2; M_g^2=0)$ and thus is defined by the expansion (5.6) in this limit, so it is
\begin{equation}
\Pi^r(q^2) = \Pi(q^2) - \Pi(0) = q^2 \Pi'(0) + q^4 \Pi''(0) + q^6 \Pi'''(0) + ... \ .
\end{equation}
The gluon wave function MP renormalization constant $Z_3(0)$ is given by the relations (5.2) and (5.5) at $M_g^2=0$, namely

\begin{equation}
 Z_3(0)=Z_3(M^2_g=0) = {\tilde Z}_3(M^2_g=0) = {1 \over 1 + \Pi(0)},
\end{equation}
i.e., the mass gap MP renormalization constant coincides with the gluon wave function MP renormalizazion constant, as it should be in the massless case.
Using these relations and definitions (5.9), from eq.~(6.3) one obtains

\begin{equation}
D_{\mu\nu}(q) = { - i \over q^2 \left[ 1 + \Pi(q^2)\right] } T_{\mu\nu}(q)
- i {\lambda}^{-1} L_{\mu\nu} (q) { 1 \over q^2},
\end{equation}
where the initial invariant function now is given by the Taylor expansion as follows:

\begin{equation}
\Pi(q^2) = \sum_{n=0}^{\infty} (q^2)^n \Pi^{(n)}(0),
\end{equation}
in agreement with the expansion (5.3), as expected, and where $\Pi^{(n)}(0)$ denotes its corresponding derivatives at the point $q^2=0$.
Eq.~(6.6) with the invariant function defined as the convergent Taylor expansion (6.7) around small $q^2$ is nothing else but eq.~(3.21), denoted as  $D^{PT}_{\mu\nu}(q)$. It has the invariant function $\Pi^{(s)}_t(q^2) = \Pi(q^2)$ regular at zero, and thus defined just by the expansion (6.7).

Neglecting now the contribution from the gluon self-energy, i.e., putting formally $\Pi(q^2) = 0$ in eq.~(6.6), for the free massless
vector boson propagator one obtains eq.~(3.3), namely
\begin{equation}
D^0_{\mu\nu}(q) = - i T_{\mu\nu}{ 1 \over q^2} - i \lambda^{-1} {q_{\mu}q_{\nu} \over q^2} { 1 \over q^2},
\end{equation}
so that from eqs.~(6.6)-(6.8) one arrives at
\begin{equation}
q_{\mu}q_{\nu} D_{\mu\nu}(q) =  q_{\mu}q_{\nu} D^0_{\mu\nu}(q) = - i {\lambda}^{-1}.
\end{equation}
Note, in order to get from eq.~(6.6) its renormalized version, one has to multiply it by the MP renormalization constant
(6.5), in accordance with the relations (5.9). Then its gauge-fixing parameter becomes different from that of the free massless gluon propagator,
again in accordance with the relations (5.9) and eq.~(6.3). Also, it is clear that eq.~(6.8) coincides with eq.~(5.12) putting
there  $M_g^2=0$, as it needs be. The PT full gluon (6.6)  and free massless gluon (6.8) propagators have the same behavior at small gluon momenta, namely
$\sim (q^2)^{-1}$. Such IR singularity is called the PT singularity in the $q^2 \rightarrow 0$ limit. Evidently, such gluon propagators
describe the propagation of massless gluons in the QCD vacuum.

It is instructive to investigate the structure of the full gluon propagator with the gluon pole mass (5.8) in the PT $q^2 \rightarrow \infty$ limit
but $M^2_g$ is finite. So that suppressing the term  $M^2_g/q^2$ in this limit in eq.~(5.8), one arrives at

\begin{equation}
D^R_{\mu\nu}(q) = { - i {\tilde Z}^{-1}_3 \over q^2 [ Z_3^{-1} + \tilde {\Pi}(q^2)]} T_{\mu\nu}(q)
-  {i \tilde {\lambda}^{-1} \over q^2} L_{\mu\nu}(q),
\end{equation}
where the expansion (5.6) now looks like

\begin{equation}
{\tilde \Pi}(q^2) = q^2 \Pi'(M^2_g) + (q^2)^2 \Pi''(M^2_g) + (q^2)^3 \Pi'''(M^2_g) + ... .
\end{equation}
Taking now in account expressions (5.2) and (5.5) as well as the definitions (5.9) and doing some algebra, one finally obtains

\begin{equation}
D_{\mu\nu}(q) = { - i  \over q^2 [ 1 + \Pi(q^2)]} T_{\mu\nu}(q) -  {i \lambda^{-1} \over q^2} L_{\mu\nu}(q),
\end{equation}
and the initial invariant function now is given by the Taylor expansion as follows:

\begin{equation}
\Pi(q^2) = \sum_{n=0}^{\infty} (q^2)^n \Pi^{(n)}(M^2_g)
\end{equation}
in agreement with the expansion (5.3) in the PT limit at finite $M^2_g$, as expected, and where $\Pi^{(n)}(M^2_g)$ denotes its corresponding derivatives at the point $q^2=M^2_g$. Contrary to the convergent Taylor expansion (6.7), the expansion (6.13) is a formal one since it demonstrates an essential singularity
at large $q^2$. So that in this case the initial invariant function becomes some function having the essential singularity in the PT limit, i.e.,
$\Pi(q^2) = F(q^2)$ at $q^2 \rightarrow \infty$. Then the corresponding effective charge looks like $d(q^2) = { 1 / 1 + F(q^2)}, \quad q^2 \rightarrow \infty$.
The behavior of any meromorphic function near its essential singularity is governed by the Picard theorem formulated in~\cite{59} (and see any textbook on the theory of functions of complex variable). It tells us that such kind of function in the close neighborhood of its essential singularity can be replaced by
the constant $Z$, which value depends only on how precisely the essential singularity is achieving. In our case the gluon momentum can go to infinity
by the two different ways, namely as a loop variable or as the free particle momentum (i.e., not a loop variable). Then the effective charge becomes simply     constant $d(q^2) = { 1 / 1 + Z}, \ q^2 \rightarrow \infty$. For example, if the gluon momentum goes to infinity as a free particle variable, we can put $Z=0$
in order to get the free gluon propagator or if it goes to infinity as s loop variable, than we can put $Z=c$, where $c$ is some constant, so that
the effective charge gets a finite renormalization. However, this all is not important, but the important observation is that in any cases the dependence
on the gluon pole mass $M^2_g$ vanishes in the massive gluon propagator. It seems reasonable conclusion, since why indeed such defined mass
should survive the PT limit, i.e., even remaining finite it cannot determine the structure of the vacuum far away from the mass-shell.
In~\cite{59} it has been explicitly shown that in the AF regime (i.e., far away from the mass-shell) the scale of the non-trivial PT dynamics
in the YM vacuum is determined by the different from $M^2_g$, indeed, the mass scale parameter $\Delta^2_{PT}$. Dynamically it is generated from the tadpole term as well, but after performing the corresponding renormalization program it can be identified with $\Lambda^2_{YM}$.

Concluding, let us make in advance a few remarks. The above-discussed effect takes place if and only if the gauge-fixing parameter is finite in the longitudinal component of the full gluon propagator (5.8). Only this determines its renormalizable behavior in the PT limit, while being the off-mass-shell
object. But its structure on the mass-shell $q^2=M^2_g$ shows some important gauge and dynamical peculiarities of the massive gluon fields (see next sections).

\section{Canonical gauge  ${\tilde\lambda}^{-1} = \infty$ }

Before going to investigate the most important issue of the massive gluon fields structure on the mass-shell $q^2=M^2_g$,
it is instructive to study one particular case. It is in close connection with the gauge structure of the massive gluon propagator (5.8)
on the mass-shell (see next section). Within the above-mentioned EGD formalism there exists one special gauge which deserves separate
consideration, namely ${\tilde\lambda}^{-1} = \lambda^{-1} = \infty$. Then eq.~(5.8) becomes
\begin{equation}
D^R_{\mu\nu}(q) = { - i Z_3 {\tilde Z}^{-1}_3 \over (q^2 - M^2_g)[ 1 + Z_3 {\tilde \Pi}(q^2)]} T_{\mu\nu}(q)
+ i{q_{\mu}q_{\nu} \over q^2} {1 \over M_g^2},
\end{equation}
and shown in Fig. 3. The corresponding ST identity now looks like
\begin{equation}
q_{\mu}q_{\nu} D^R_{\mu\nu}(q) = - i \xi^R(q^2) = i {q^2 \over M_g^2}.
\end{equation}

Neglecting now the contribution from the regular part of the full gluon self-energy in accordance with the relations (5.11), for the free massive
counterpart in this gauge, one obtains
\begin{equation}
D^0_{\mu\nu}(q; M^2_g) = {- i \over (q^2 - M^2_g)} \left( g_{\mu\nu} - {q_{\mu}q_{\nu} \over M^2_g} \right) \sim \textrm{const.}  \quad \textrm{in the $q^2 \rightarrow \infty $ limit},
\end{equation}
which coincides with eq.~(5.12) in the $\lambda^{-1} = \infty$ limit, as expected, and the same ST identity, namely
\begin{equation}
q_{\mu}q_{\nu} D^0_{\mu\nu}(q; M^2_g)= -i \xi_0(q^2)  = i {q^2 \over M_g^2}.
\end{equation}

From the expressions (7.1)-(7.2) and especially clearly from eqs.~(7.3)-(7.4) it follows that massive gluon propagator in this gauge has
unrenormalizable behavior in the PT $q^2 \rightarrow \infty$ limit, and the $M^2_g=0$ limit does not exist as well, see Fig. 3.
In the quantum electroweak gauge theory such asymptotic, coming from the longitudinal component
of the propagators for massive vector particles $Z, W^+, W^-$ is not a problem. Due to the conserved currents in this theory, longitudinal components
of their propagators do not contribute to the $S$-matrix elements, describing this or that physical process/quantity.
In QCD such conserved currents do not exist, and moreover, the longitudinal components interact with each other~\cite{4}, so the unrenormalizable behavior
of the massive gluon propagator possesses a serious problem in this gauge. It has been called as canonical one
in~\cite{29}. Let us emphasize that we have derived the ST identity for the massive vector particle (7.2) within our GGI/EGI formalism, i.e., not addressing to its standard derivation, which, for example, has been performed for the massive YM effective theory in~\cite{44}.
In~\cite{45} the case of the $M^2_g=0$ limit has been investigated in some details as well.

For further discussion, let us formulate the self-consistent condition for the gauge choice in QCD as follows:

\begin{equation}
{{\tilde\lambda}^{-1} q^2 \over q^2 -  {\tilde\lambda}^{-1} M_g^2} = {a q^2 \over q^2 -  a M_g^2},
\end{equation}
i.e., the left-hand-side of this equation is present by the generalized gauge expression (5.10), while its right-hand-side presents the same expression
when the gauge is already chosen. In other words, we are checking wether the GGD formalism is compatible with its EGD one and vice versa, so that its aim is
to derive a relation (not an identity) involving the gauge-fixing parameter.
If $a$ is any finite number, then from the self-consistently relation (7.5) it is easy to derive that $ {\tilde\lambda}^{-1} = a$, indeed.
At the same time, if $a = \infty$ then the relation (7.5) becomes

\begin{equation}
{{\tilde\lambda}^{-1} q^2 \over q^2 -  {\tilde\lambda}^{-1} M_g^2} = - {q^2 \over M_g^2},
\end{equation}
which is only satisfied at $q^2=0$, i.e., no any condition imposed on the gauge-fixing parameter. These simple arguments point out that
something is really wrong with the canonical gauge ${\tilde\lambda}^{-1} = \infty$ in QCD.

All this shows the general inconsistency of the canonical gauge in QCD within the mass gap approach to its ground state.
In other words, only finite gauges ${\tilde\lambda}^{-1}$ are compatible with the mass gap approach to QCD.
It formulates the interaction and gauge
of the massive full gluon propagator (5.8), associated with its transverse and longitudinal components, respectively, in such a way that the theory
remains renormalizable. The effective charge, associated with its transverse component, is finite in the PT $q^2 \rightarrow \infty$.
The invariant function, associated with its
longitudinal component, is a known function of $q^2$ and  $M^2_g$. It explicitly demonstrates the equivalence between the PT $q^2 \rightarrow \infty$
and $M^2_g=0$ limits and vice versa, which is important for the renormalizability. Thus, the massive gauge particle propagator (the corresponding Green's function) is at least quadratically convergent, i.e., behaves as $\sim (1/q^2)$ at $q^2 \rightarrow \infty$. In other words, our approach renders the YM massive theory renormalizable at infinity. Also, it provides the smooth transition $\xi \rightarrow \xi_0$ in the PT or $M^2_g=0$ limits.
However, the resulting (i.e., renormalized) massive gluon propagator in the canonical gauge ${\tilde\lambda}^{-1} = \infty$ (Fig. 3) makes the theory unrenormalizable both off-mass-shell, eq.~(7.1), and on the mass-shell $q^2=M^2_g$, eq.~(7.3).
Thus, the above-mentioned equivalence is broken down by this gauge. Therefore, the canonical gauge should
be abandoned, as not satisfying to the PT behavior at infinity and having no smooth $M^2_g=0$ limit, both requested by the mass gap approach to QCD.
Let us note that the $\lambda^{-1}={\tilde\lambda}^{-1} = \infty$ limit in QCD seems to us too artificial and formal, like the formal $M^2=0$ one, indeed.

Concluding this section by a few remarks, trying to explain from the physical point of view the mathematical inconsistency of the canonical
gauge in QCD, let us note that this is not the whole story yet, see section below.

\begin{figure}[h]
\begin{center}
\includegraphics[width=15.0truecm]{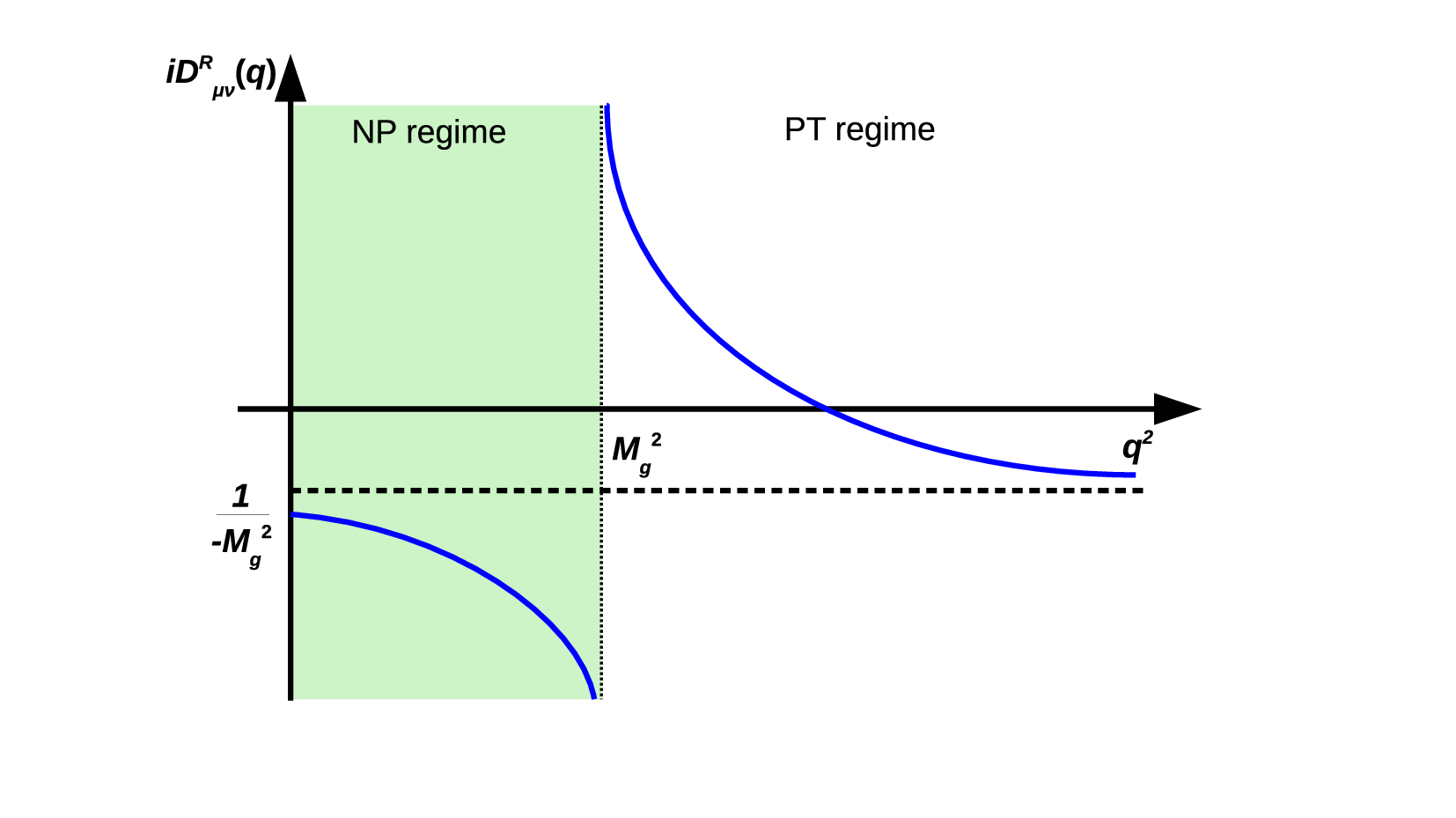}
\caption{The massive gluon propagator (7.1) in the canonical gauge ${\tilde\lambda}^{-1} = \infty$ as a function of $q^2$.
It breaks down the PT $q^2 \rightarrow \infty$ limit in comparison with Fig. 2.}
\label{fig:3}
\end{center}
\end{figure}

\section{The mass-shell $q^2=M^2_g$ }

The one of the important issues of a mass dynamical generation in QCD is to be considered here. For this purpose, let
us discuss in detail the structure of the massive gluon propagator at finite ${\tilde\lambda}^{-1}$
on the mass-shell $q^2=M^2_g$ or simply mass-shell in what follows. It is closely related to the free massive state.
Within the GGD formalism, i.e., not specifying the gauge-fixing parameter ${\tilde\lambda}^{-1}$ first, the massive full
gluon propagator (5.8) on the mass-shell becomes
\begin{equation}
D^R_{\mu\nu}(q) = { - i Z_3  {\tilde Z}^{-1}_3 \over (q^2 - M^2_g)} \left( g_{\mu\nu} - { q_{\mu}q_{\nu} \over M_g^2} \right)
- i \tilde{\epsilon} { q_{\mu}q_{\nu} \over M_g^4},
\end{equation}
because of the first of the relations (5.7) and where

\begin{equation}
\tilde{\epsilon} = { {\tilde\lambda}^{-1} \over (1 - {\tilde\lambda}^{-1})} =  { 1 \over ({\tilde\lambda} - 1)}, \quad \tilde\lambda^{-1} \neq 1.
\end{equation}

However, it is not difficult to show that eq.~(8.1) is equivalent to
\begin{equation}
D^R_{\mu\nu}(q) = { - i Z_3  {\tilde Z}^{-1}_3 \over (q^2 - M^2_g)} \left( g_{\mu\nu} - { q_{\mu}q_{\nu} \over M_g^2} \right)
\end{equation}
on the mass-shell at any $\tilde{\epsilon}$, so that for its free massive counterpart one arrives at

\begin{equation}
D^0_{\mu\nu}(q; M^2_g) = { - i \over (q^2 - M^2_g)} \left( g_{\mu\nu} - { q_{\mu}q_{\nu} \over M_g^2} \right),
\end{equation}
since in this case the combination $Z_3  {\tilde Z}^{-1}_3 = 1$ in eq.~(8.3).
It is in agreement with eq.~(5.12) on the mass-shell, which does not depend on the gauge
choice at all as well.
At the same time, eq.~(8.4) coincides with the expression (5.13) on the mass-shell which has been derived in unitary gauge ${\tilde\lambda}^{-1} = 0$.
In its turn it is agreed with eq.~(8.1) at $\tilde{\epsilon}= 0$ or, equivalently, ${\tilde\lambda}^{-1} = 0$, see eq.~(8.2).
It is worth emphasizing that eq.~(8.4) coincides with the free massive gluon propagator (7.3) in the canonical gauge  ${\tilde\lambda}^{-1} = \infty$.
Thus, one concludes that the free massive gluon propagator (5.12) on the mass-shell, which equivalent to eq.~(8.4),  has unacceptable
behavior in the PT $q^2 \rightarrow \infty$ limit, explicitly seen in Fig. 3, and such behavior does not depend on any gauge choice.

It is instructive now to present the massive full gluon propagator in the canonical gauge (7.1) at the mass-shell $q^2=M^2_g$, so it is

\begin{equation}
D^R_{\mu\nu}(q) = { - i Z_3 {\tilde Z}^{-1}_3 \over (q^2 - M^2_g)} \left( g_{\mu\nu} - { q_{\mu}q_{\nu} \over M_g^2} \right)
+ i{q_{\mu}q_{\nu} \over M^4},
\end{equation}
since reminding ${\tilde \Pi}(q^2=M^2_g) = 0$, as it follows from the relations (5.7). However, it is easy to show that this expression
on the mass-shell is equivalent to

\begin{equation}
D^R_{\mu\nu}(q) = { - i Z_3 {\tilde Z}^{-1}_3 \over (q^2 - M^2_g)} \left( g_{\mu\nu} - { q_{\mu}q_{\nu} \over M_g^2} \right),
\end{equation}
and its free massive counterpart becomes

\begin{equation}
D^0_{\mu\nu}(q) = { - i \over (q^2 - M^2_g)} \left( g_{\mu\nu} - { q_{\mu}q_{\nu} \over M_g^2} \right),
\end{equation}
since also reminding that in this case $Z_3 {\tilde Z}^{-1}_3 = 1$. However, comparing the system of eqs.~(8.3)-(8.4) with those of (8.6)-(8.7),
one can conclude that they are the same and do not depend on any gauge choice, as it needs be for any gluon propagator being on the mass-shell.
The self-consistency condition for the full massive gluon propagator can be formulated by equating eq.~(8.1) to eq.~(8.5) since both are being
on the mass-shell, indeed. Then one obtains

\begin{equation}
\tilde{\epsilon} = { {\tilde\lambda}^{-1} \over (1 - {\tilde\lambda}^{-1})} = - 1,
\end{equation}
with the same result as for the consideration off-mass-shell gluon propagators in the previous section.
The self-consistency condition is not satisfied for the full massive gluon propagator on the mass-shell as well ($-1 \neq 0$).

However, there exists only one exception from the general picture described above. Within the EGD formalism let us begin with the
t'Hooft\,--\,Feynman gauge ${\tilde\lambda}^{-1}=1$ in eq.~(5.8). Then going to the mass-shell and setting  the combination $Z_3  {\tilde Z}^{-1}_3 =1$ again, one obtains

\begin{equation}
D^0_{\mu\nu}(q; M^2_g) = { - i \over (q^2 - M^2_g)} g_{\mu\nu}
\end{equation}
for its free massive counterpart and in complete agreement with eq.~(5.14). It has the renormalizable behavior at $q^2 \rightarrow \infty$ because of the automatic cancellation of the longitudinal components on the mass-shell in this particular gauge. Just this cancellation provides smooth
$M^2_g=0$ limit in eq.~(8.9). This principal difference between all other finite gauges and the t'Hooft\,--\,Feynman gauge explicitly
shown in eq.~(8.9), underlines that the distinction between GGD and EGD formalisms, introduced in Section IV within the mass gap approach
to QCD, is not only a mere convention, but it makes sense, indeed (see also the test on the gauge choice's self-consistency
condition, described above and in the previous section). At the same, it is difficult to assign any physical meaning to eq.~(8.9),
even being on the mass-shell, since it is explicitly gauge-dependent, and it is not transverse.

Our general conclusion is that the massive gluon propagator cannot be the mass-shell object within the mass gap approach to QCD, and thus it cannot appear in the physical spectrum (confinement of massive gluons). On the contrary, staying off-mass-shell, the massive gluon propagators keeps save the above-mentioned equivalence between $q^2 \rightarrow \infty$ and $M^2_g=0$ limits in its longitudinal component at any finite gauge, required by the mass gap approach to QCD. Thus, the massive gluons may only exist in the QCD ground state or inside hadrons, treated as its color-singlet massive excitations.

Concluding, the regularization prescription $i0^+$ has been omitted in the massive gluon propagators.
They are substantially modified due to the response of the NP QCD vacuum. This prescription is designated for and can be applied
to the theories with perturbative vacua \cite{8,63}.
In the massless perturbative gluon propagators in Section VI it has been omitted for simplicity.
Also, it is clear that we are working in the covariant gauges in order to avoid the peculiarities of the non-covariant (axial) gauges \cite{8}.

\section{Summary}

One of the important conceptual problems in theoretical physics is the origin of a mass~\cite{47}, and hence the existence of the mass gap itself~\cite{9}. Our findings provide new insights into its dynamical generation at the fundamental quark-gluon level.
We have explicitly shown that the initial exact SU(3) color gauge symmetry of the QCD Lagrangian is not a symmetry of its ground state (vacuum) from the dynamical point of view, without mentioning its well-known topological complexities at the classical level~\cite{48} (instantons, monopoles, etc). Quite possible that just due to our claim that the symmetries of the QCD Lagrangian and its ground state do not coincide, QCD is a self-consistent quantum field gauge theory. Therefore it needs no extra degrees of freedom in order to dynamically generate a mass.

Within our approach the gauge symmetry of the ground state has not been broken down by hand. By the design it is there from the very beginning.
It has been broken by the presence of the tadpole term in the dynamical and gauge structures of the full gluon SD equation, which describes the propagation of gluons in the QCD ground state. The tadpole term, having the dimensions of mass squared, is generated by the self-interaction of massless gluon modes, but
the point-like quartic gluon vertex is only involved at the skeleton loop level, see Fig. 1 and discussions in~\cite{8,9,64}. However, the exact gauge symmetry requires its omission along with other mass scale parameters having the dimensions of mass squared. So it plays no role in the preservation of this symmetry in the QCD vacuum.
In other words, the tadpole term is explicitly present in the QCD vacuum, but its true role has been hidden by the exact gauge symmetry. The derived splintering expressions (3.37) makes it possible to reveal the tadpole term, which renormalized version we call the mass gap, as the dynamical source of this symmetry breakdown in the vacuum of QCD. In this way, we extend the concept of the mass gap to be also accounted for the QCD ground state.
Symbolically it is possible to say that the mass gap has been 'closed' in the room and the splintering is a key to open it, and thus to disclose the real role of the mass gap in the true dynamical and gauge structures of the QCD ground state. So that the splintering procedure permits to explicitly show up the mass gap in the full gluon SD equation and thus in the full gluon propagator as well, while respecting the corresponding ST identities for it and its free counterpart.

All this allows to formulate novel NP approach to QCD and its ground state in Section IV. We call it as the mass gap approach.
The general structure of the full gluon propagator, expressed in terms of the regularized quantities, has been derived and shown in eq.~(4.5).
Its longitudinal component is exactly defined as a function of $q^2$ and $M^2$, shown in eq.~(4.6). Its transverse component is
defined up to the invariant function $\Pi(q^2; D)$, which is regular at zero and may have the PT logarithmic divergences only: otherwise remaining arbitrary.
The full gluon invariant function (4.2) is nothing else but the transcendental equation.
If one formally ignores the tadpole contribution $M^2$, then we are left with the
PT massless full gluon propagator (3.21), and ignoring further the contribution from its invariant function, one arrives at the free
massless gluon propagator (3.3). The renormalization of the PT full gluon propagator is well-known procedure, see for example~\cite{3,4,5}, and it is not
our problem here, as emphasized above. We have demonstrated that the exact gauge symmetry of the QCD Lagrangian is dynamically broken down by the explicit presence of the mass gap in its ground state. This violation is to be accompanied by the gauge-changing phenomenon, $\xi \neq \xi_0$. We have also explain how the exact gauge symmetry of the QCD Lagrangian might be preserved in its vacuum, which leads to the relation $\xi = \xi_0$ (Section III).

In close connection with the mass gap generation and gauge symmetry breakdown occurs the problem of the formulation of the corresponding renormalization theory. The tadpole term being the dynamical source of the gauge symmetry breakdown can not be renormalized within the PT method, since it is quadratically divergent constant, but regularized one. Despite the gauge symmetries of the QCD Lagrangian and its ground state are not the same,
nevertheless, the renormalizability of QCD/YM theory is not affected, i.e., it remains renormalizable within the mass gap approach to the QCD ground state.
The renormalization program has been performed in two steps. The PT unrenormalizable quadratic UV divergences, associated with the skeleton loop diagrams
and included into the quark and gluon constants $\Delta^2_q$ and $\Delta^2_g$, respectively, have been removed from the theory by the
developing self-consistent and gauge-invariant subtraction scheme (complemented by the corresponding transverse conditions) in Section III.
The second step was to formulate the NP MP renormalization program for the tadpole term (3.43) itself, so that the renormalized mass gap becomes
the exactly defined gluon pole mass. This program performed in Section V renders theory to the renormalizable logarithmic divergences only.
All the non-physical parameters (such as regulating the quadratic UV divergences or arbitrary subtraction points, etc.) have been incorporated into the corresponding NP MP renormalization constants of the mass gap and the gluon wave function.
The dependence on these constants is not so important in the purely YM theory, but their correct definitions (5.2) and (5.5) in this theory are important.
In this way, the general expression for the renormalized full massive gluon propagator (5.8) was derived. It has been written down in a newly-derived
generalized gauge (5.10). The characteristic feature of the NP MP program for the massive gluon propagator is that the above-mentioned NP MP renormlization
constants (5.2) and (5.5) are different, and they coincide for the massless case only, see eq.~(6.5).

The important issue of the renormalization beyond the PT method within the mass gap approach to QCD needs a few clarifying remarks in addition.
For the renormalization of the PT QCD it is essential to remove from the theory all the quadratically divergent scale parameters
since it fails to renormalize them. The exact gauge symmetry makes it possible to achieve this goal. For the NP approach the presence
of the quadratically divergent scale parameters (which violet the exact gauge symmetry) is not a problem. In this paper we have explicitly shown
how the NP MP renormalization program has to be performed for the regular massive gluon modes, while the intrinsically NP (INP) MP renormalization program
for the severely singular massless gluon modes has been performed in~\cite{59}. Briefly speaking, the mass scale parameters
(the quadratically divergent from above or severely singular from below) being an essential problems for the PT QCD, are not the problems
for our general approach, i.e., being unrenormalizable in the PT QCD such quantities can be renormalized in the NP/INP QCD.
In other words, in the PT QCD the exact gauge symmetry can not be broken in its ground state. In the NP QCD and in its INP phase
the exact gauge symmetry has to be broken, otherwise impossible to explain color confinement and scale violation in AF regime~\cite{59}.
The color confinement and AF, being the experimental facts, should be considered as the 'boundary conditions' for any solution to QCD
at the fundamental quark-gluon level.
There is no doubt that the dynamical and gauge structures of the PT QCD and NP/INP QCD ground states are substantially different.
Let us emphasize that when we are speaking (above and below) that the QCD Lagrangian's gauge symmetry is broken down or violated or not
the same, etc. as in its ground state, we mean that the terms (having the dimensions of mass squared) which cannot be present in the
Lagrangian may still present in the ground state. Moreover, they can survive the corresponding renormalization program if it goes beyond
the PT technics, i.e., such kind of programs (depending on the above-mentioned solutions) are to be essentially NP. Precisely in this
sense the renormalizability of QCD/YM as a theory of strong interactions is to be understood

In the presence of the mass gap the coupling constant $g^2$ plays no any role. This is also evidence of the 'dimensional transmutation',
$g^2 \rightarrow M^2_g$~\cite{2,60,61}, which occurs whenever a massless theory acquires mass dynamically. It is a general feature of spontaneously symmetry breaking in field theories. Therefore, we distinguish between the PT and NP QCD by the explicit presence of the mass scale parameter -- the mass gap -- in the latter one, and not by the magnitude of the coupling constant. Fig. 2 directly indicates the existence of the phase transition
at $q^2= M^2_g$ between the NP and PT regimes in the QCD vacuum. There is no smooth transition between them when the gluon momentum squared goes from zero to infinity and vice versa. In other words, this is a phase transition with an essential discontinuity at the point $q^2= M^2_g$, since the left- and
right-sided limits are the corresponding infinities.
The existence of the massive solution shows the general possibility for a massless vector particle to acquire mass dynamically in the
vacuum without any extra degrees of freedom. Let us remind once more an important observation that any deviation
of the full gluon propagator from the free one requires the presence of the mass scale parameter. We consider the tadpole term as such mass scale parameter. That is why there is no need  within our approach to introduce to QCD the mass scale parameter by hand or to extract it from the theory by some other way.

The important issue of the peculiarities of the canonical gauge  $\tilde{\lambda}^{-1} = \lambda^{-1} = \infty$ for the full massive
gluon propagator has been investigated in Section VII.
Within the mass gap approach to QCD the canonical gauge should be abandoned in this theory
because of its two unacceptable features. Firstly, it makes the gauge particle Green functions unrenormalizable in the PT limit.
Secondly, it breaks down the equivalence between the PT $q^2 \rightarrow \infty$  and the $M^2_g=0$ (and vice versa) limits in the longitudinal component
of the massive full gluon propagator, and thus breaks the correct transition $\xi \rightarrow \xi_0$ in the above-mentioned limits (see Appendix A as well).
This equivalence is requested by the mass gap approach to NP QCD in order to ensure its renormalizability.
Only the finite gauge-fixing parameters can guarantee this.
The important role of the ST identity (5.10) in the formulation of the self-consistency condition
is to be underlined. It is based on the comparison between the GGD and EGD formalisms introduced in Section IV.
Due to this condition the canonical gauge is shown to be inconsistent in the massive YM theory, and thus the application of the GGD/EGD formalism
is useful in this theory.

The behavior of the massive full gluon propagator (5.8) on the mass-shell $q^2=M^2_g$ was the subject of our investigation in Section VIII.
It has been explicitly shown that no any physical meaning can be assigned to the gauge particle propagators with exactly defined gluon pole masses. Being on the mass-shell, these propagators very much look like the massive gluon propagator in the canonical gauge, apart from
the t' Hooft - Feynman gauge. The gauge-independent structure on the mass-shell does not exist for the massive gluon propagators within our approach.
Moreover, being on the mass-shell, the full gluon propagator does not satisfy the transition $\xi \rightarrow \xi_0$ in the PT $q^2 \rightarrow \infty$  and  $M^2_g=0$ limits (and vice versa). The self-consistency condition, formulating within the ST identity, shows the inconsistency of the mass-shell structure
for the full massive gluon propagator. So that the GGD/EGD formalism, on which the self-consistency condition is based, turned out to be effective in this case as well.
However, staying off-mass-shell, the massive gluon propagator satisfies the equivalence between the PT $q^2 \rightarrow \infty$  and $M^2 \rightarrow 0$ limits
and thus keeps the above-mentioned transition valid, which are requested by the mass gap approach. The massive gluons may exist only in the QCD ground state and inside hadrons, treated as its colour-singlet excitations. In the QCD vacuum and in hadrons they present new gluon degrees of freedom different from the gluon excitations with the effective gluon masses. The renormalization properties of the ST identity provides the finite results beyond the mass-shell
for the full massive gluon propagator. Such kind off-mass-shell objects cannot appear as physical states at large distances (confinement of massive
gluons with exactly defined gluon pole masses). In any case, the massive gluon propagator is strongly suppressed at large distances ($q^2 \rightarrow 0$)
in comparison with its confining singular counterpart~\cite{59}.

The existence of the mass gap has been studied within lattice QCD in recent paper~\cite{51}. However, its mass gap cannot be identified with the gluon pole mass, but rather being the mass which appear in the singular solution~\cite{59}.
The pioneering work~\cite{52}, where the gluon mass has been introduced into the continuum QCD, is to be acknowledged.
The discussion of other approaches and models for
a possible existence of the gluon mass, for example such as Curci\,--\,Ferrari~\cite{53}, Gribov\,--\,Zwanziger~\cite{54,55}, the effective gluon
mass function~\cite{31,56}, lattice simulations~\cite{32} (and see references therein), dual QCD~\cite{57}, LCO formalism to calculate
the gluon pole mass~\cite{58}, etc., is to be performed independently elsewhere.

However, one important observation can be made even at this stage. The general expression for the full massive gluon propagator derived
without any use of the truncations/approximations/assumptions for the corresponding invariant function, associated with its transverse component, and special gauge choice, associated with its longitudinal component, has been present in eq.~(4.5). Therefore, its renormalized version's regular behaviour at zero gluon momentum (5.16)  is gauge-invariant and exact, see also Fig. 2. Moreover, when the mass gap can be omitted in the PT theory limit, the corresponding gluon propagator (3.21) may have only the same singular structure as the free gluon propagator has at $q^2 \rightarrow 0$ limit (see the derivation in Section 6).
In other words, we do not change the NL transcendental character of the initial sum of the skeleton loop integrals in the gluon SD eq.~(2.1), and shown
in Fig. 1. Their NL and transcendentality has been saved and present in the gluon SD eq.~(4.5) simply in other form.
So that we did not loss any peace of the useful information on the massive gluon field configurations within our solution, investigated here.
We consider this as a big advantage of the mass gap approach to QCD.
At the same time, in many papers cited in recent review~\cite{62} the explicit derivations of the above-mentioned invariant function (which might be called the effective gluon mass function by omitting the mass gap contribution in eq.~(4.5)) have been inevitably made on the basis of the different truncations/approximations/assumptions/ansatzs and schemes as well as the special gauges choice, mainly unitary (Landau) gauge.
So that the initial transcendentality of the gluon invariant function has been affected from the very beginning due to a such kind of the simplifications.

\hspace{2mm}

In short, the resume of our main results is: the discovery of the true dynamical and gauge structures of the QCD vacuum
explains confinement of massive gluon states. The crucial role in this discovery
has been played by the extension of the JW mass gap concept~\cite{9}, introduced within the Hamiltonian framework of QCD, to be also accounted
for its ground state. Our mass gap coincides with the JW's one by properties, but not by definition. Within the massive solution every excitation of the vacuum has energy at least $\Delta = M_g$, indeed. At the same time, how to prove that the YM Hamiltonian (H) has no spectrum in the interval $(0, M_g)$
is a completely different problem and remains thus beyond the scope of the present investigation. One of our main goals here was to prove the existence
of the mass gap in the QCD ground state and its survival after the NP renormalization program has been performed.
The challenges, formulated in Section I, have been addressed and solved by the mass gap approach to QCD ground state.
By solution we mean first of all the formulation of the corresponding NP renormalization program maintaining the renormalazibility of QCD
in the explicit presence of the quadratically divergent constants, and thus remaining unrenormalizable by the PT technics.
The conceptual problem of confinement
of massless gluons and thus the general problem of confinement of all types of color gluons has been addressed and solved by our approach in~\cite{59},
where the solutions of the other three conceptual problems mentioned above have been given as well. Also, the mass gap approach allows to determine the asymptotical and analytical properties of the gauge particles (gluons) Euclidean and Minkowskian renormalized full Green's functions. They describe
the propagation of the severely singular massless gluons investigated in~\cite{59} and the regular massive gluons investigated in
the present work, respectively.

Whether the massive gluon propagator can be useful for the solution of the quark confinement problem is a perspective subject for the
further work.

Finally we note, unlike the ghost and tadpole terms the gauge-fixing term of the QCD Lagrangian does not explicitly contribute to the full gluon self-energy, that is why it has been left out of our consideration here.

\begin{acknowledgments}

The authors are grateful to P. Forg\'{a}cs, J. Nyiri, T.S. Bir\'{o}, M. Vas\'{u}th, Gy. Kluge
for useful suggestions, remarks, discussions and help.
The work was supported by the Hungarian National Research, Development and Innovation Office (NKFIH) under the contract numbers OTKA K135515, K123815 and NKFIH 2019-2.1.11-T\'ET-2019-00078, 2019-2.1.11-T\'ET-2019-00050, the Wigner GPU Laboratory and THOR Cost Action CA15213.

\end{acknowledgments}

\appendix

\section{ The non-splintering procedure }

The splintering procedure in respect with how to explicitly show that the exact gauge symmetry/invariance
in the QCD ground state is already broken from the very beginning has been described in Section III. The only other possible way
to do this is presented in this Appendix. As we already know from the relations (3.5) and (3.8) in the general case
\begin{equation}
q_{\rho} q_{\sigma} \Pi_{\rho\sigma} (q; D) \neq 0,  \quad \quad      q_{\rho} q_{\sigma} \Pi^{(s)}_{\rho\sigma} (q; D) \neq 0,
\end{equation}
i.e., both transverse conditions are not satisfied, which means there is no the splintering (3.37) at all. However, in the similar way exploited in Section III, let us begin with the corresponding subtraction
\begin{equation}
\Pi^{(s)}_{\rho\sigma}(q; D)= \Pi_{\rho\sigma}(q; D) - \Pi_{\rho\sigma}(0; D)= \Pi_{\rho\sigma}(q; D) - \delta_{\rho\sigma} \Delta^2(D),
\quad \textrm{with} \quad \Pi^{(s)}_{\rho\sigma}(0; D) =0,
\end{equation}
where by $\Delta^2(D)$ we denote the sum which appears in eq.~(3.7), namely $\Delta^2(D) = \Delta^2_q + \Delta^2_g(D) + \Delta^2_t(D)$.
The general decompositions of the full gluon self-energy and its subtracted counterpart into the independent tensor structures are given in
the relations (3.9), namely
\begin{eqnarray}
\Pi_{\rho\sigma}(q; D) &=&  - T_{\rho\sigma}(q) q^2 \Pi_t(q^2; D) + q_{\rho} q_{\sigma} \Pi_l(q^2; D), \nonumber\\
\Pi^{(s)}_{\rho\sigma}(q; D) &=& - T_{\rho\sigma}(q) q^2 \Pi^{(s)}_t(q^2; D) + q_{\rho} q_{\sigma} \Pi^{(s)}_l(q^2; D).
\end{eqnarray}
The both invariant functions $\Pi^{(s)}_t(q^2; D)$ and $\Pi^{(s)}_l(q^2; D)$ cannot have power-type singularities (or, equivalently, pole-type ones) at small $q^2$, since $\Pi^{(s)}_{\rho\sigma}(0; D) =0$ by definition in eq.~(A2): otherwise they remain arbitrary. On account of the subtraction (A2) and the relations (A3), one obtains
\begin{eqnarray}
\Pi_t (q^2; D) &=&  \Pi^{(s)}_t(q^2; D) - {\Delta^2(D) \over q^2}, \nonumber\\
\Pi_l(q^2; D) &=&   \Pi^{(s)}_l(q^2; D) + {\Delta^2(D) \over q^2},
\end{eqnarray}
then the full gluon self-energy becomes
\begin{equation}
\Pi_{\rho\sigma}(q; D) = - T_{\rho\sigma}(q) \left[ q^2 \Pi^{(s)}_t(q^2; D) - \Delta^2(D) \right]
+ L_{\rho\sigma} \left[ q^2 \Pi^{(s)}_l(q^2; D) + \Delta^2(D) \right].
\end{equation}
Substituting eq.~(A5) into the initial gluon SD eq.~(2.1), one arrives at

\begin{equation}
D_{\mu\nu}(q) = D^0_{\mu\nu}(q) - D^0_{\mu\rho}(q)i T_{\rho\sigma}(q) \left[ q^2 \Pi^{s}_t(q^2; D) - \Delta^2(D) \right] D_{\sigma\nu}(q)
+ D^0_{\mu\rho}(q)i L_{\rho\sigma}(q) [ q^2 \Pi^{(s)}_l(q^2; D) + \Delta^2(D)] D_{\sigma\nu}(q).
\end{equation}
Contracting both sides of this equation with $q_{\mu}$ and $q_{\nu}$ and doing some algebra using the decompositions (3.2)
and (3.3), one finally gets

\begin{equation}
\xi \equiv \xi(q^2) = { \xi_0 q^2 \over q^2 - \xi_0 [\Delta^2(D) + q^2 \Pi^{(s)}_l(q^2; D)] },
\end{equation}
instead of eq.~(4.4). Let us remind that in this expression $\Delta^2(D) = \Delta^2_q + \Delta^2_g(D) + \Delta^2_t(D)$, while in
the solution (4.4) it is the tadpole constant itself, i.e.,  $\Delta^2_t(D)$.
If both transverse conditions (A1) were satisfied, i.e., equal zero, then $\Delta^2_q = \Delta^2_g(D) = \Delta^2_t(D) =0$, and
we would have reproduced the correct limit of the exact gauge symmetry.

The last expression (A7) in the PT $q^2 \rightarrow \infty$ limit becomes

\begin{equation}
\xi \equiv \xi(q^2) = { \xi_0  \over 1 - \xi_0 \Pi^{(s)}_l(q^2; D) },
\end{equation}
which is equivalent to the formal $\Delta^2(D) = 0$ one. However, as we already know, the arbitrary invariant function $\Pi^{(s)}_l(q^2; D)$ may have logarithmic divergences
at $q^2 \rightarrow \infty$. So that $\xi(q^2) \rightarrow - 0$ in this limit, while in the generalized gauge (4.4) it goes to
$\xi(q^2) \rightarrow \xi_0$, which is only one correct. In order to get the correct limit from the general expression (A7) one needs to formally put
$\Pi^{(s)}_l(q^2; D) =0$ there. Then from the relations (A3) and (A4) one finally arrives at the splintering relations (3.37), indeed.
This is one more strong argument in favor of our formalism, developed in Section III, in order to prove that the gauge symmetries of the QCD
Lagrangian and its ground state are not the same. Remind also that the second of the transverse relations (3.37) requires
$\Delta^2_q = \Delta^2_g(D) =0$, while $\Delta^2(D) = \Delta^2_t(D)$ remains finite.

On the other hand, combining eq.~(A5) and the general transverse condition (3.5), one obtains

\begin{equation}
[\Delta^2(D) + q^2 \Pi^{(s)}_l(q^2; D)] = \left( {\xi_0-\xi \over \xi \xi_0} \right) q^2,
\end{equation}
and substituting it back to the previous eq.~(A7), one arrives at the identities $\xi= \xi$ as well as $\xi_0 = \xi_0$.
This is in complete agreement with the identities which come from the initial gluon SD eq.~(3.12) contracted with $q_{\mu}$ and $q_{\nu}$.
We consider these identities as a test on the correctness and uniqueness of the mass gap approach to QCD.
They confirm that the gauge choice by the relation (3.32) and the equivalent solution (4.4) have been justified.
Only the expression (4.4) has a correct PT $q^2 \rightarrow \infty$ limit, which is equivalent to the formal $\Delta^2(D) = 0$ one, as it was described
in Section IV.
The unique way to correctly show up the mass scale parameter in the full gluon propagator
is the splintering expressions (3.37), explicitly based on the tadpole term as the mass gap.

\section{Euclidean metric for comparison to Lattice calculation}

Let us begin with pointing out that the gluon pole mass is different from the effective gluon mass.
The former one is exactly defined by the relation (5.1) in a gauge-invariant way, while the latter one is to be extracted from
the effective gluon mass function, see
for example~\cite{31}. In this connection, it is worth noting that our initial expression is the transcendental one, namely (4.2).
Our investigation is of purely theoretical nature, so we cannot fix the numerical value of
$M^2_g$. However, this is possible to do within lattice QCD simulations \cite{32}, but within our approach.
For this purpose, let us write down the full massive gluon propagator (5.8) in Euclidean metric by making the substitution $q^2 \rightarrow - q^2$
in (5.8), so that $q^2=q^2_0 + \bf q^2$ becomes from now on, and thus the mass-shell is defined as $q^2= - M^2_g$. Then one obtains
\begin{equation}
D^R_{\mu\nu}(q) = { i Z_3  {\tilde Z}^{-1}_3 \over (q^2 + M^2_g)[ 1 + Z_3 {\tilde \Pi}(q^2)]} T_{\mu\nu}(q) + i { q_{\mu}q_{\nu} \over q^2} { {\tilde\lambda}^{-1} \over (q^2 + {\tilde\lambda}^{-1} M_g^2)},
\end{equation}
where now $T_{\mu\nu}(q) = \delta_{\mu\nu} - L_{\mu\nu}(q)$. The expansion for the invariant function ${\tilde \Pi}(q^2)$ then looks like
\begin{equation}
{\tilde \Pi}(q^2) = (q^2 + M^2_g) \left[ \Pi'(-M^2_g) + q^2 \Pi''(-M^2_g) + q^2(q^2 + M^2_g) \Pi'''(-M^2_g) + ... \right],
\end{equation}
while for the corresponding NP MP renormalization constants one arrives at
\begin{equation}
{\tilde Z}_3^{-1}= [ 1  +  \Pi(-M^2_g)], \quad Z_3^{-1} = {\tilde Z}_3^{-1} - M^2_g \Pi'(-M^2_g).
\end{equation}
It is instructive to repeat the derivation of eq.~(B1) at $q^2=0$ in this metric in detail as it has been done in Section V.
So first one obtains
\begin{equation}
D^R_{\mu\nu}(0) = { i  {\tilde Z}^{-1}_3 \over M^2_g[ Z_3^{-1} + {\tilde \Pi}(0)]} T_{\mu\nu}(q) + i { q_{\mu}q_{\nu} \over q^2} { 1 \over M_g^2}.
\end{equation}
Taking now into account eq.~(B3) and eq.~(B2) at $q^2=0$, one finally obtains
\begin{equation}
i D^R_{\mu\nu}(0) = - { 1 \over M^2_g } T_{\mu\nu}(q) -  { q_{\mu}q_{\nu} \over q^2} { 1 \over M_g^2} = - { 1 \over M^2_g}  \delta_{\mu\nu},
\end{equation}
coinciding completely with eq.~(5.16), as it needs be (see Figs. 2 and 3).

It will be useful to introduce
the following notation as $\Pi^{(n)}(-M^2_g) = a_n, \quad n = 0,1,2,3$, where $a_n$ are the arbitrary constants of the corresponding dimensions.
Then eq.~(B1) after doing some simple algebra will be expressed in terms of these constants only, namely
\begin{equation}
iD^R_{\mu\nu}(q) = - { (1 + a_0) T_{\mu\nu}(q) \over (q^2 + M^2_g)[ (1 + a_0) + a_1 q^2 + (q^2 + M^2_g)q^2(a_2 + a_3 (q^2 + M^2_g) + ...)]}
 - { q_{\mu}q_{\nu} \over q^2} { {\tilde\lambda}^{-1} \over (q^2 + {\tilde\lambda}^{-1} M_g^2)}.
\end{equation}
The higher order terms ($a_4$ and higher) in its denominator can be
suppressed in the NP region $ -M^2_g \leq q^2 \leq 0$ we are mainly interested in. We are not interested in the PT tail of the massive
gluon propagator, i.e., in the perturbative region $ q^2 \leq -M^2_g$. For this reason, it is much more convenient for lattice simulations to use its free massive counterpart as follows:

\begin{equation}
iD^0_{\mu\nu}(q) =  - { 1 \over (q^2 + M^2_g)} \left[ \delta_{\mu\nu} - (1 - \lambda^{-1})
{q_{\mu}q_{\nu} \over (q^2 + \lambda^{-1} M^2_g)} \right],
\end{equation}
and especially this expression in the t'Hooft\,--\,Feynman gauge $\lambda^{-1}=1$, which is
\begin{equation}
iD^0_{\mu\nu}(q) = - { 1 \over (q^2 + M^2_g)}\delta_{\mu\nu}.
\end{equation}
Compare these equations with eq.~(5.12) and eq.~(5.14) in Minkowski metric. However, let us remind that the values at zero $q^2=0$ and at
$q^2= - M^2_g$ are exactly and gauge-invariantly defined. Also our estimation \cite{8} of the scale of the NP dynamics in QCD was about $(0.5-0.6)$ GeV.
The pole mass of a single gluon is expected to be of the same order of magnitude. Note that the value of $m_g \approx 550$ MeV has been calculated in \cite{32} (and see references therein for other determinations).

\subsection*{Remarks on the gluon pole mass}

The massive gluons with non-zero pole masses, which are exactly defined in (5.1), may exist in the
QCD ground state and inside hadrons along with their massless counterparts.
We call such a solution for the full massive gluon propagator as NP massive (since the gluon pole mass is of purely NP dynamical origin, as described above).
Due to its asymptotic properties, the gluon pole mass contribution is to be neglected in the PT $q^2 \rightarrow \infty$ limit.
This means that in the experiments at high energies involving the strongly-interacted particles, it is difficult to fix the gluon pole mass $M_g^2$.
On the other hand, in the low-energy experiments it cannot be detected as well, since the gluon pole mass solution is to be totally suppressed by its singular
counterpart~\cite{59} at low-energies (large distances). Apparently, "experimentally" the gluon pole mass can be fixed by lattice simulations, as
formulated just above. The dynamically generated gluon pole mass looks like very similar to the current mass of a free quark.
The principle difference is that now we know how it may appear in the full gluon propagator.
The dynamical source of the current quark mass is still remains unknown, though its term is compatible with the $SU(3)$ color gauge invariance of the QCD Lagrangian, while the gluon pole mass term not.

There are no any doubts that inside hadrons and nuclei quarks can interact with each other by exchange of not only massless gluons but their massive counterparts as well. The interaction between heavy quarks due to the exchange of massive gluons is described by the Yukawa-type potential $V(r) \sim (1/r) \exp(-M_gr)$, though it is not confining quarks, it is strong but short-range. This means that one can 'see' the massive gluons inside
the strongly-interacting particles (hadrons) only. But one cannot 'see' them inside hadrons at short distances ($q^2 \rightarrow \infty$), as pointed out above at the end of Section VI.
The massive gluons will contribute to the quarks effective masses (properly defined) via the quark SD equation, making them much more different
from the current quark masses, mentioned above. This knowledge may substantially improve our understanding of the dynamical structure of
the QCD ground state as well as hadron and nuclei internal structures and their properties, such as masses~\cite{33}, nucleon
and glue spins, etc.~\cite{34,35}. The existence of massive gluons can provide more information about the ordinary nuclear matter in the interior
of compact (neutron) stars and their merger~\cite{36}. The gluons with exactly defined pole masses may also play important role in the creating of the different phases of QCD Matter at high temperature and density~\cite{37,38,39,40}. They will provide new gluon degrees of freedom, but
different from the quasi-particles which also show up as effective poles (depending on the temperature) in the gluon propagators, calculated by the thermal lattice QCD~\cite{41}. Apparently, the massive gluons should be somehow included into the YM NP equation of states as well, e.g. such as derived in~\cite{42}. One can easily imagine the glueballs as bound states of the two or three gluons with pole masses and not only consisting of the gluons with effective gluon masses or massless gluons~\cite{43}.

Concluding our discussion, let us note that at $M^2_g=0$ the above-mentioned Yukawa-type potential becomes of the Coulomb-type, namely $V(r) \sim (1/r)$.
Thus it describes the interaction between quarks due to exchange of the PT (6.6) and free (6.8) massless gluons.


{}


\begin{thebibliography}{}
\bibitem{1}
   H. Fritrsch, M. Gell-Mann, H. Leutwyler, Phys. Lett. B, 47 (1973) 365.
\bibitem{2}
   W. Marciano, H. Pagels, Phys. Rep. C, 36 (1978) 137.
\bibitem{3}
   M.E. Peskin, D.V. Schroeder, An Introduction to Quantum Field Theory (Addison-Wesley, 1995).
\bibitem{4}
   C. Itzykson, J.-B. Zuber, Quantum Field Theory (McGraw-Hill Book Company, 1984).
\bibitem{5}
   T. Muta, Foundations of QCD (Word Scientific, 1987).
\bibitem{6}
   N. Brambilla et al., Eur. Phys. J. C, 37 (2014) 2981. 
\bibitem{7}
   A.S. Kronfeld, C. Quigg, Am. J. Phys., 78 (2010) 1081. 
\bibitem{8}
   V. Gogokhia, G.G. Barnaf\"oldi, The Mass Gap and its Applications (World Scientific, 2013).
\bibitem{9}
   A. Jaffe, E. Witten, Yang\,--\,Mills Existence and Mass Gap, \\
   $http://www.claymath.org/prize-problems/, \
   http://www.arthurjaffe.com$ \ .
\bibitem{10}
   E.G. Eichtein, F.L. Feinberg, Phys. Rev. D, 10 (1974) 3254.
\bibitem{11}
   M. Baker, C. Lee, Phys. Rev. D, 15 (1977) 2201.
\bibitem{12}
   U. Bar-Gadda, Nucl. Phys. B, 163 (1980) 312.
\bibitem{13}
   R. Alkofer, L. von Smekal, Phys. Rep. C, 353 (2001) 281.
\bibitem{14}
   C.D. Roberts, A.G. Williams, Prog. Part. Nucl. Phys., 33 (1994) 477.
\bibitem{15}
   P. Maris, C.D. Roberts, Int. J. Mod. Phys. E, 12 (2003) 297.
\bibitem{16}
   J. C. Taylor, Nucl. Phys. B, 33 (1971) 436.
\bibitem{17}
   A. A. Slavnov, Sov. Jour. Theor. Math. Phys., 10 (1972) 153.
\bibitem{18}
   G. 't Hooft, Nucl. Phys. B, 33 (1971) 173.
\bibitem{19}
   S.K. Kim, M. Baker, Nucl. Phys. B, 164 (1980) 152.
\bibitem{20}
   S.-H.H. Tye, E. Tomboulis, E.C. Poggio, Phys. Rev. D, 11 (1975) 2839.
\bibitem{21}
   P. Pascual, R. Tarrach, Nucl. Phys. B, 174 (1980) 123.
\bibitem{22}
   B.W. Lee, Phys. Rev. D, 9 (1974) 933.
\bibitem{23}
   S. Aoki et al., Eur. Phys. J. C, (2017) 77:112. 
\bibitem{24}
   V.N. Gribov, J. Nyiri, Quantum Electrodynamics (Cambridge University Press, 2001).
\bibitem{25}
   V. Gogokhia, Phys. Lett. B, 618 (2005) 103.
\bibitem{26}
   G. 't Hooft, M. Veltman, Nucl. Phys. B, 44 (1972) 189.
\bibitem{27}
   G. 't Hooft, Nucl. Phys. B, 35 (1971) 167.
\bibitem{28}
   T.D. Lee, C.N. Yang, Phys. Rev., 128 (1962) 885.
\bibitem{29}
   K. Fujikawa, B.W. Lee, A.I. Sanda, Phys. Rev. D, 6 (1972) 2923.
\bibitem{30}
   B.W. Lee, J. Zinn-Justin, Phys. Rev. D, 7 (1973) 1049.
\bibitem{31}
   B. Holdom, Phys. Lett. B, 728 (2014) 467.
\bibitem{32}
   A. Cucchieri, D. Dudal, T. Mendes, N. Vandersickel, Phys. Rev. D, 85 (2012) 094513.
\bibitem{33}
   M. Tanabashi, {\it et al.}, Particle Data Group, Phys. Rev. D, 98 (2018) 030001.
\bibitem{34}
   C.A. Aidala, S.D. Bass, D. Hasch, G.K. Mallot, Rev. Mod. Phys., 85 (2013) 655. 
\bibitem{35}
   Yi-Bo Yang et al., Phys. Rev. Lett., 118 (2017) 102001-1.
\bibitem{36}
   A. Jakov\'ac, G.G. Barnaf\"oldi, P. P\'osfay, Phys. Rev. C, 97 (2018) 025803.
\bibitem{37}
   U. Heinz et al., \ arXiv:1501.06477 [nucl-th].
\bibitem{38}
   {\it Quark Matter 2014}, Proc. of the XXIV Inter. Conf. on Ultra-Relativistic Nucleus-Nucleus Collisions, edited by:
   P. Braun-Munziger, B. Friman, J. Stachel, Nucl. Phys. A, 931 (2014) 1
\bibitem{39}
   Melting Hadrons, Boiling Quarks, From Hagedorn Temperature to Ultra-Relativistic heavy-Ion Collisions
   at CERN, edited by J. Rafelski (Springer Open, 2015).
\bibitem{40}
   P. Petreczky, J. Phys. G: Nucl. Part. Phys., 39 (2012) 093002. 
\bibitem{41}
   P. Petreczky, F. Karsch, E. Laermann, S. Stickan, I. Wetzorke, Nucl.Phys.Proc.Suppl. 106 (2002) 513-515. 
\bibitem{42}
   V. Gogokhia, M. Vasuth, J. Phys. G: Nucl. Part. Phys., 37 (2010) 075015.
\bibitem{43}
   C. Bernard, Phys. Lett. B, 108 (1982) 431.
\bibitem{44}
   J. Gegelia, G. Japaridze, Mod. Phys. Lett. A, 27 (2012) 1250128. 
\bibitem{45}
   J. Gegelia and U.-G. Meisner, Eur. Phys. J. A, 50 (2014) 174. 
\bibitem{46}
   Y.L. Dokshitzer, D.E. Kharzeev, Ann. Nucl. Part. Sci., 54 (2004) 487. 
\bibitem{47}
   F. Wilczek, Phys. Today, Nov. issue (1999) 11; ibid, Jan. issue (2000) 13.
\bibitem{48}
   V. Rubakov, Classical Theory of Gauge Fields
   (Prinseton University Press, 2002).
\bibitem{49}
   V. Gogokhia, G.G. Barnaf\"oldi, Int. J. Mod. Phys. A, 31 (2016) 1645027:
\bibitem{50}
   V. Gogohia, Phys. Lett. B, 584 (2004) 225.
\bibitem{51}
   R. Vilela-Mendes, Int. J. Mod. Phys. A, 32 (2017) 1750016.
\bibitem{52}
   J.M. Cornwall, Phys. Rev. D, 26 (1982) 1453.
\bibitem{53}
   G. Curci, R. Ferrari, Nuovo Cim., A35 (1976) 1.
\bibitem{54}
   V.N. Gribov, Nucl. Phys. B, 139 1978) 1.
\bibitem{55}
   D. Zwanziger, Nucl. Phys. B, 321 1989) 591.
\bibitem{56}
   A.C. Aguilar, D. Binosi, J. Papavassiliou, Front. Phys., (11)2 (2016) 111203.
\bibitem{57}
   M. Baker, J.S. Ball, F. Zachariasen, Phys. Rep. 209 (1991) 73.
\bibitem{58}
   J.A. Gracey, Eur. Phys. J. C, 39 (2005) 61.
\bibitem{59}
   V. Gogokhia, G.G. Barnaf\"oldi, \ arXiv:2012.15337 \ [hep-th, hep-ph, math-ph, nucl-th].
\bibitem{60}
   S. Coleman, E. Weinberg, Phys. Rev. D, 7 (1973) 1888.
\bibitem{61}
   D.J. Gross, A. Neveu, Phys. Rev. D, 10 (1974) 3235.
\bibitem{62}
   M.Q. Huber, Phys. Rept., 879 (2020) 1-92, \ arXiv:1808.05227 [hep-ph].
\bibitem{63}
   Y.L. Dokshitzer, D.E. Kharzeev, Ann. Nucl. Part. Sci., 54 (2004) 487. 
\bibitem{64}
   R. Feynman, Nucl. Phys. B, 188 (1981) 479.
\bibitem{65}
   S. Weinberg, The quantum theory of fields (Cambridge University Press, 1995).
\end{thebibliography}
\end{document}